\documentclass{article} 
\usepackage{nips14submit_e,times}
\usepackage{moreverb,url}
\usepackage[pdftex]{graphicx}
\usepackage{amsmath}
\usepackage{amssymb}
\usepackage[justification=centering]{caption}
\usepackage{graphicx}

\title{A Fast and Efficient Near-Lossless Image Compression using Zipper Transformation}

\author{
Babajide O. Ayinde \\
Electrical and Computer Engineering\\
University of Louisville\\
Louisville, KY 40218 \\
\texttt{babajide.ayinde@louisville.edu} \\
}

\nipsfinalcopy 

\begin{document}

\maketitle

\begin{abstract}
Near-lossless image compression-decompression scheme is proposed in this paper using Zipper Transformation (ZT) and inverse zipper transformation (iZT). The proposed ZT exploits the conjugate symmetry property of Discrete Fourier Transformation (DFT). The proposed transformation is implemented using two different configurations: the interlacing and concatenating ZT. In order to quantify the efficacy of the proposed transformation, we benchmark with Discrete Cosine Transformation (DCT) and Fast Walsh Hadamard Transformation (FWHT) in terms of lossless compression capability and computational cost. Numerical simulations show that ZT-based compression algorithm is near-lossless, compresses better, and offers faster implementation than both DCT and FWHT. Also, interlacing and concatenating ZT are shown to yield similar results in most of the test cases considered.
\end{abstract}


\section{Introduction}
Lossless compression scheme is important in many critical application domains such as biomedical image storage, art images, remote sensing, security and defense, just to mention a few \cite{brunello2003lossless,baligar2003high,li2002image}. In teleradiological applications, one of the key players for  effective diagnosis as well as the treatment is the quality of the image \cite{sriraam20113}. Since in most medical applications, data is often acquired and stored digitally, therefore, the demand for storage and transmission of multidimensional medical images has sprung up some interests in this knowledge domain \cite{srikanth2005contextual,xiong2003lossy}. Medical images are often very large in size and number, and any means to compress them can lead to reduction of storage cost and increased transmission throughput since data are often stored on servers and relayed to clients on demand. For instance, a 256 MB is required for a 512 $\times$ 512 $\times$ 512 16-bit computed tomography (CT) image data \cite{nguyen2011efficient}. To visualize and analyze this data is mostly computationally inefficient and daunting. The primary objective of compression in medical applications is reducing large volume of data to be stored and transmitted while preserving crucial diagnostic information \cite{nguyen2011efficient}. Even though the cost of storage and transmission of digital signal has plummeted, however, the demand for lossless compression of medical images is increasing \cite{miaou2009lossless}.\\
\indent
From a general viewpoint, image compression can be broadly divided into two main categories: lossy and lossless compression. Lossy compression deals with compression schemes that have tolerance for some certain amount of error, that is, the compressed and the decompressed images may not be identical. In contrast, lossless compression encodes all the information from the original image and therefore, the decompressed image is identical to the original image. Lossy compression is sometimes tolerated in some medical applications as long as the necessary and required diagnostic information is preserved in the decompressed image \cite{sriraam20113,xiong2003lossy,tai2000adaptive,sunder2005performance}.\\
\indent
In medical imaging, lossy compression can sometimes achieve a minute compression before a good percentage of information is dropped. More compression can be achieved if some visible losses can be tolerated for clinical task purposes. There are still a lot of controversies as to what the real life applications of lossy compressions are, especially in the medical domain. One other approach to lossy compression is the machine learning approach where images are encoded with sparse feature representations using neural networks \cite{Hinton2006reducing,wang2012folded,hashim2016optimal}.\\
\indent
Generally speaking, lossless compression can be categorized into three broad categories, namely: predictive scheme with statistical modeling, transform based coding and finally, dictionary based coding. The predictive deals with using statistical method to evaluate the differences between pixels and their neighbors, and performed context modeling before coding. Whereas in transform based compression, pixel are transformed using frequency or wavelet transformation before modeling and coding. Dictionary based compression is the third category and it deals with replacing strings of symbols with shorter codes. It must be noted that dictionary based schemes are widely used for text compression \cite{egorov2015performance}. An example of a dictionary based algorithm is the well known ZIP package. Other dictionary based compression algorithms for image data are the Lempel-Ziv-Welsh (LZW), Portable Network Graphics (PNG), Graphics Interchange Format, and so on.\\
 \indent
A good number of image data compression paradigms have been investigated in recent past. The least squares adaptive prediction scheme was proposed in \cite{li2001edge} for lossless compression  of natural images and it was shown that the novel scheme improves the computational complexity with negligible performance trade-off. Lossless compression based on adaptive spectral band re-ordering and adaptive backward previous closest neighbors (PCN) algorithms was also proposed in \cite{zhang2007efficient,wang2005lossless} for hyperspectral images, and it was shown that the compression performance was greatly enhanced with the implementation of both the re-ordering of spectral band. The problem of compression in medical applications has also been studied in \cite{clunie2000lossless} where different types of compression standards on grayscale medical images were compared by highlighting the pros and cons of various compression methods.\\
\indent
In \cite{brunello2003lossless}, problems of video compression were addressed taking cognizance of the temporary spectral information. Also in \cite{memon1996lossless}, the possibility of using 3-D versions of the lossless JPEG spatial predictors was considered and the likelihood of using best predictor to encode the present frame was investigated. The spectral redundancy was also exploited by implementing the best predictor from one spectral component to another spectral component. It can be  inferred from \cite{memon1996lossless} that pixels in a given neighborhood are concurrent in adjoining color bands. To improve on this, a different predictor for interband correlation was proposed in \cite{memon1997interband}. In \cite{yang2000contex}, a simple context predictive scheme was also proposed where either intraframe or interframe coding is selected based on temporal and spatial variations, and the prediction of the current pixel was then computed. A good number of predictors were considered taking cognizance of the spatial redundancy.\\
 \indent
In addition, a variation of singular value decomposition (SVD) based image compression was proposed in \cite{ranade2007variation} which is an extension on the conventional SVD-based compression. Image data was first preprocessed using data independent permutation prior to computing SVD, and in the reconstruction algorithm, inverse permutation was used to post-process the output of the conventional SVD-based decompression algorithm. Many compression algorithms have been proposed for a wide range of imaging domains such as DNA microarray images \cite{hernandez2016progressive} and multimedia image data \cite{tomar2015lossless}.\\
\indent
Improving the lossless compression of images with sparse histogram was addressed in \cite{pinho2002online} and its robustness was shown on other types of images. Both lossy and lossless compression was investigated in a unified framework and a new cost effective coding strategy was proposed to enhance the coding efficiency. A motion-JPEG-LS based lossless scheme was also proposed in \cite{miaou2009lossless} and the work only explores high enough correlation between adjacent image frames in order to avoid possible coding loss and abrupt high computational cost. Wavelet-based lossy to lossless compression methods have also been addressed for ECG \cite{miaou2005wavelet} and for volumetric medical images \cite{bruylants2015wavelet}. Lossless digital audio compression scheme was addressed in \cite{hans2001lossless}.\\
\indent
The main contribution of this paper is to propose new near-lossless compression heuristics and benchmark their performance with DCT and FWHT-based compression paradigms in term of computational complexity and compression ratio. The algorithms have been implemented taking into consideration the three important figure-of-merit namely: the compression efficiency, easiness of use, and speed of computation. The design requires no complicated  parameter selection nor special digital signal processing hardware add-on. This paper considerably expands the scope of the ZT first introduced in \cite{ayinde2016lossless} by: (i) proposing new interlacing ZT (ii) showing the performance similarity between interlacing and concatenating ZT, and lastly (iii) generalizing the proposed compression scheme to N-dimensional data. The rest of the paper is organized as follows. Section 2 describes the implementation of the zipper, inverse zipper transformation; section 3 presents the N-dimensional zipper transformation and zipper-based Huffman coding and section 4 discusses the experimental designs and presents the results. Finally, conclusions are drawn in section 5.
\section{Zipper Transformation}
\begin{figure}[!htbp]
  \centering
  \includegraphics[width=1.0\linewidth]{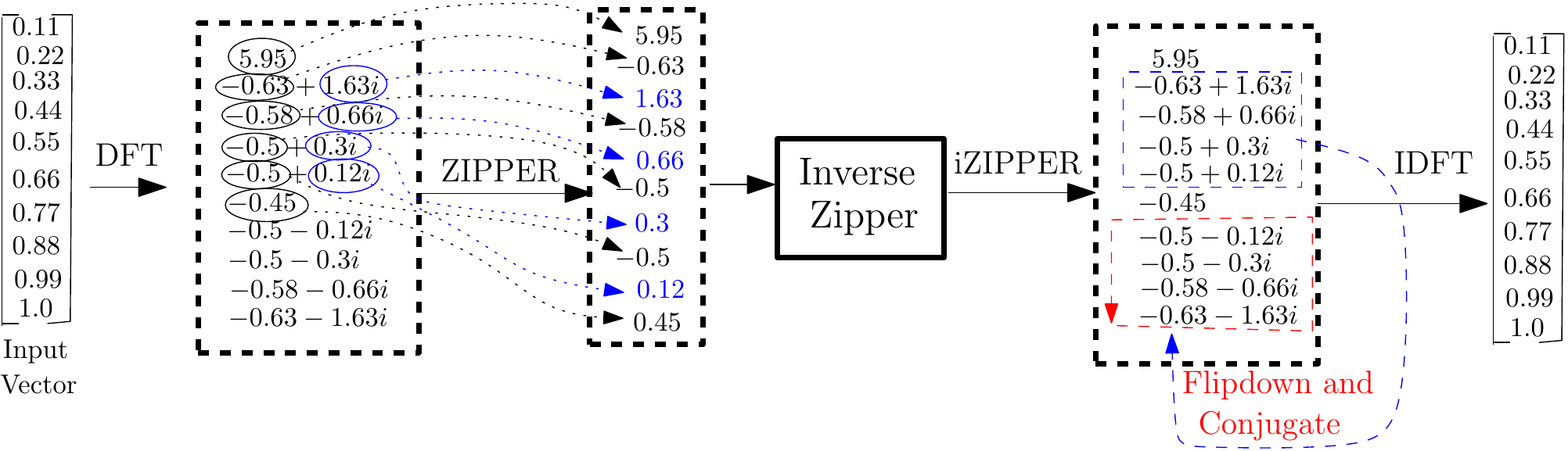}
  \captionsetup{justification=centering}
  \caption{Illustration of Interlaced-zipper and Inverse Interlace-zipper Transformation}\label{example_interlace}
\end{figure}

The main idea in this transformation is to first implement Discrete Fourier Transform (DFT) on the input array. For image data, it must be noted that DFT extracts the frequency component of each pixel. For a given set of real valued discrete samples $f(n)$, the DFT is given as:
\begin{equation} \label{MyEq1}
\begin{split}
F(k) = \sum^{N-1}_{n=0}f(n)W_N^{kn},
\end{split}
\end{equation}
where
\begin{equation} \label{MyEq2}
\begin{split}
W_N^{kn} = \exp^{-j\frac{2\pi kn}{N}} \;\;\; n = 0,...,N-1
\end{split}
\end{equation}
The inverse DFT is given by:
\begin{equation} \label{MyEq3}
\begin{split}
f(n) = \frac{1}{N}\sum^{N-1}_{k=0}F(k)W_N^{-kn}
\end{split}
\end{equation}
The output of DFT is a set of complex numbers having a property known as the conjugate symmetry. ZT exploits the conjugate symmetry property of DFT. After the deployment of DFT, the spatial representation is translated into the frequency domain representation. The complex elements of the vector in the upper half of the symmetry is extracted and examined for spectral dependencies.
\begin{figure}[!htbp]
  \centering
  \includegraphics[width=1.0\linewidth]{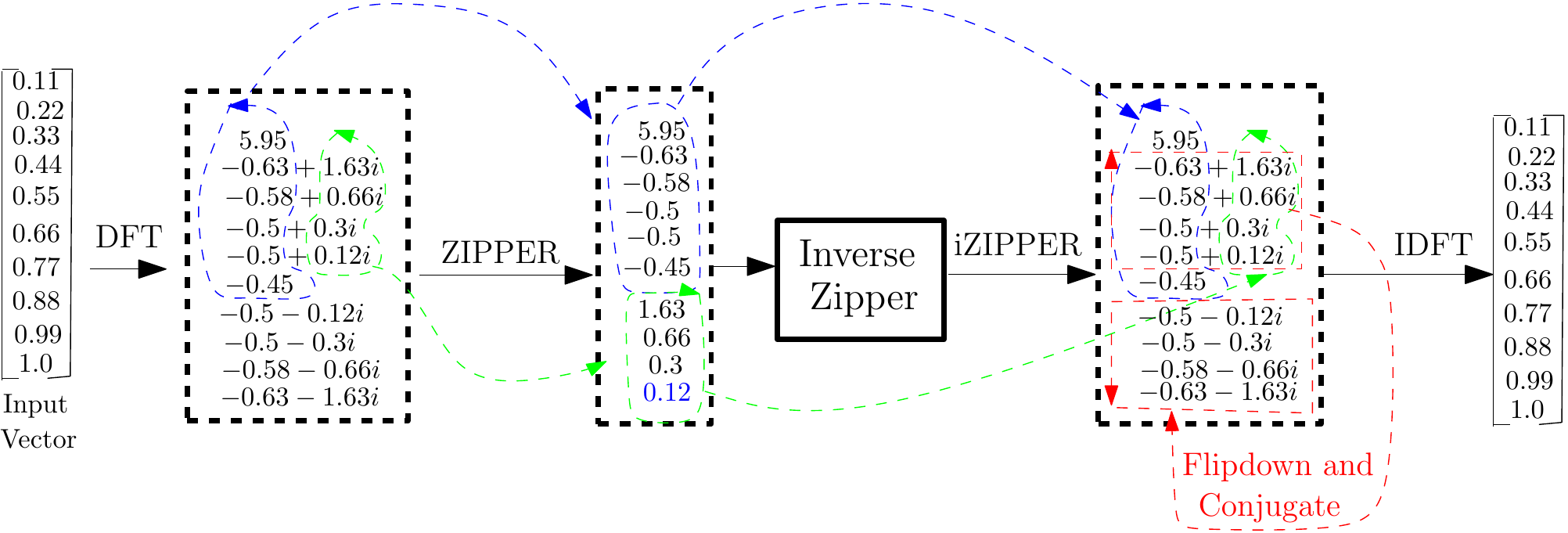}
  \captionsetup{justification=centering}
  \caption{Illustration of Zipper and Inverse Zipper Transformation}\label{example}
\end{figure}
In this paper, ZT is implemented using two different heuristics namely: the interlacing and the concatenating zipper transform. In fact, the interlacing zipping operation is abstracted from the way two separate entities can be fastened and tightened with a zipping tool. In the case of interlacing-ZT, the imaginary parts of the complex numbers in the upper half of the symmetry are stripped off and interlaced with their corresponding real counterparts. In this configuration, the structural arrangement of the imaginary parts and real part guides against arbitrary variation of pixel intensities in the spatial domain. In the concatenating ZT (or simply ZT) on the other hand, the imaginary parts of the complex numbers in the upper half of the symmetry are stripped off and concatenated with their corresponding real counterparts.\\
\indent
It is noted that this transformation is near-lossless. Also, the two-dimensional zipper transformation is implemented as a two 1D ZT in sequence by first performing a 1D ZT on columns  and then doing another 1D ZT on rows. This procedure is reversed in the inverse zipper transformation. Fig.~\ref{example} illustrates the zipper and inverse-zipper transformations and Fig.~\ref{example_interlace} illustrates the implementation of interlacing-zipper and inverse-interlacing-zipper transformations using a $10\times1$ vector. It is remarked that the number of elements in the vector before ZT stays the same after the transformation, that is, ZT preserves the dimensionality of the input data. For images, the input vector illustrated in Figs.~\ref{example} and \ref{example_interlace} is the columns or rows of the multidimensional array representing the image of interest.
\section{Zipper Transformation for N-dimensional Image Data}
\begin{figure}[!htbp]
  \centering
  \includegraphics[width=1.0\linewidth]{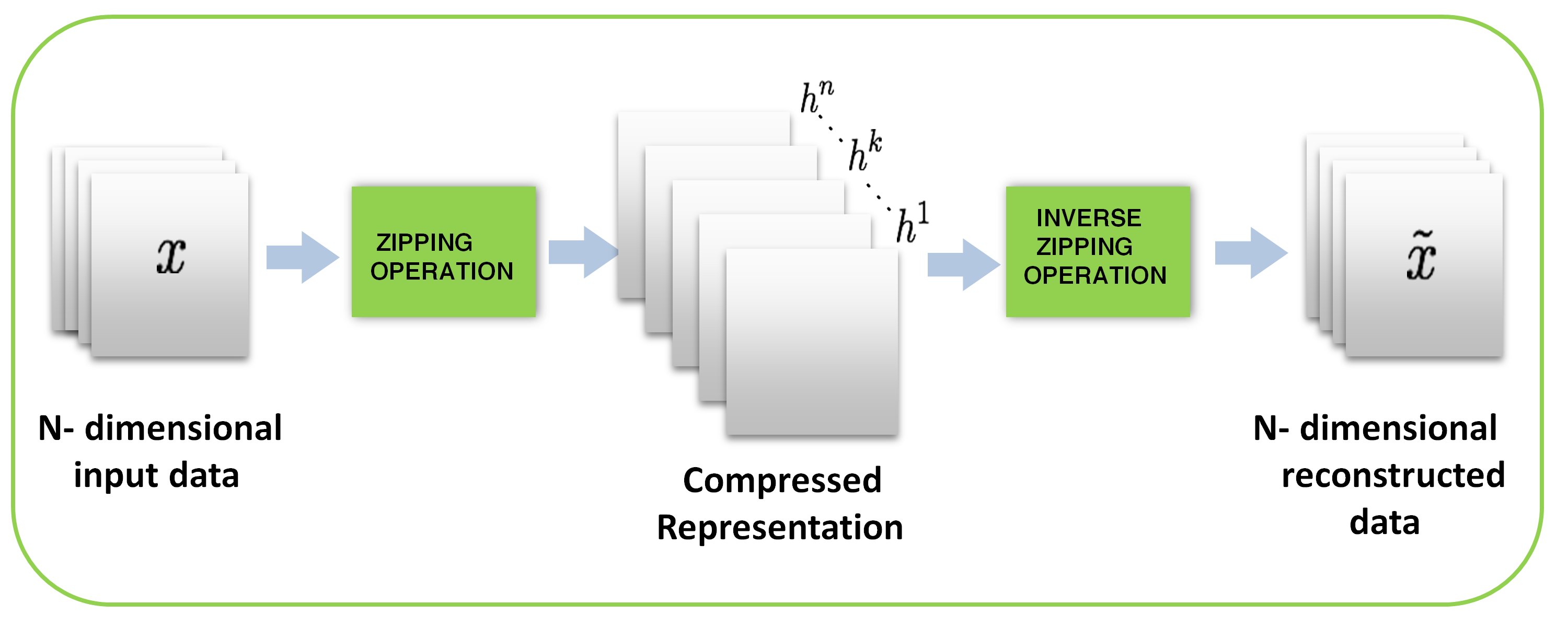}
  \captionsetup{justification=centering}
  \caption{Zipper and Inverse-zipper Transformation for N-dimensional image data}\label{nd_zipper}
\end{figure}

It many medical applications, 3-D images are more often than not preferred for analysis as it provides the flexibility of viewing the anatomical cross sections required for accurate detection of abnormalities \cite{sriraam20113}. In this section, multi-dimensional zipper transformation is proposed to compress high dimensional images such as 3-D medical images. Just like many other transformations such as wavelet transform and DCT \cite{he2003optimal,sriraam20113}, N-dimensional ZT can be realized by applying 1-D ZT along each dimension as shown in Fig.~\ref{nd_zipper}.

\subsection{Lossless Image Transformation and Coding} 
\begin{figure}[!htbp]
  \centering
  \captionsetup{justification=centering}
  \includegraphics[width=1.0\linewidth]{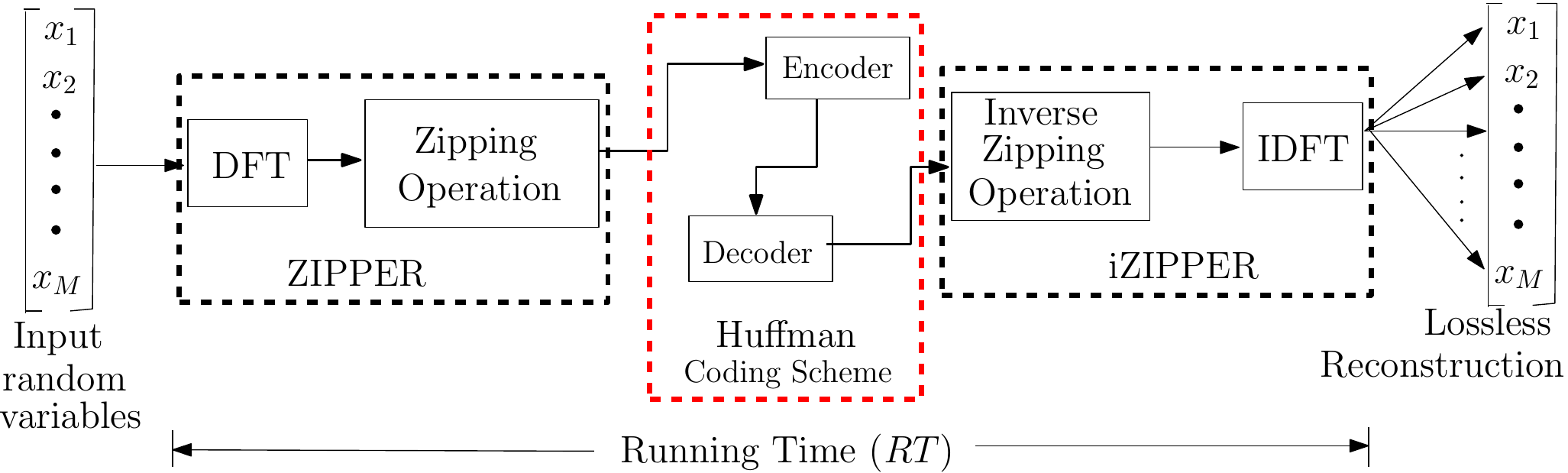}
  \caption{Lossless image compression pipeline using Zipper Transformation}\label{zipper}
\end{figure}

In this section, the pixel intensities of image data are transformed using the ZT and the output is encoded using efficient coding scheme. The basic idea of this approach is to obtain a different distribution in the transformed domain that could make symbol coding more efficient. By virtue of this transformation, the transformed image has a lower entropy, and hence, coding scheme such as Huffman coding is able to encode this stream of data with codewords of variable length. At the receiving end, the decoder can then use the encoded information and a lookup table (or dictionary) to retrieve the information back. It is worthy to know that these encoding and decoding operations must also be lossless \cite{verhack2015lossless}. It is remarked that other efficient coders such as Lempel-Ziv-Welch (LZW) algorithm can also be used in place of Huffman coding. The inverse zipper transformation can then be utilized to recover the original image with no loss incurred. The schematic of the whole process is as shown in Fig~\ref{zipper}.
\subsection{Huffman Coding}
This method was proposed in 1952 by David Huffman to compress data by reducing the amount of bits necessary to represent a string of symbols. The Huffman coding technique is a commonly used scheme for data compression because it is very simple and efficient. It uses the statistical information about the distribution of the data to be encoded. Besides, an identical coding table is used in both the encoder and the decoder. Huffman coding use codewords with variable length, and with shortest codewords for most frequently used characters. The flowchart of the Huffman coding is given below in Fig~\ref{huff}.
\begin{figure}[h]
  \centering
  \captionsetup{justification=centering}
  \includegraphics[width=8cm]{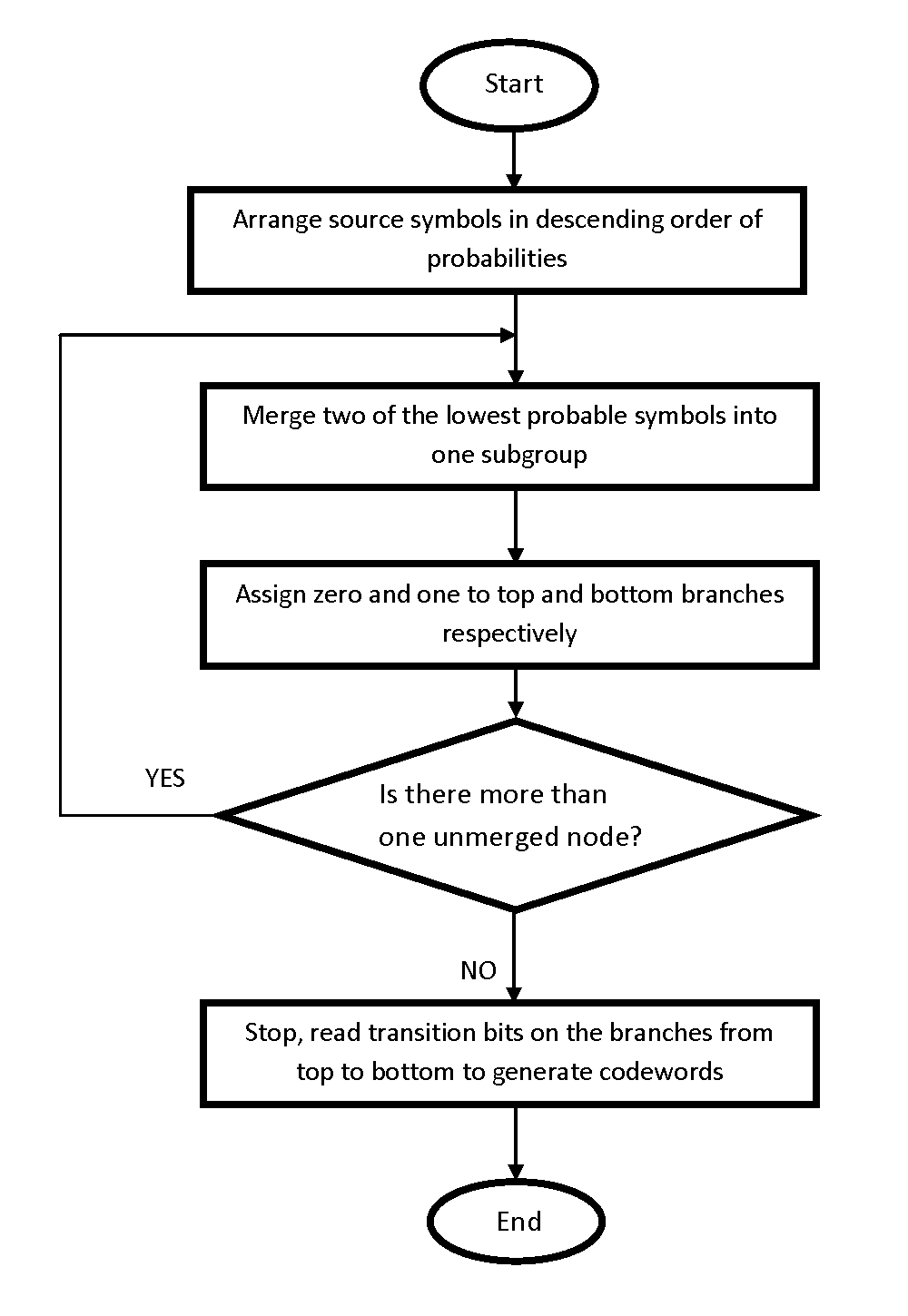}
  \caption{Huffman Coding Flowchart}\label{huff}
\end{figure}
More formally, for a discrete random variable $\textbf{X}$ with underlying finite set with $n$ elements such that $\textbf{X} = \{x_1,x_2,\cdots,x_n\}$. Given that the probability of the $i^{th}$ element $P(x_i)$ is denoted as $p_i$, then entropy of $\textbf{X}$ is calculated as:
\begin{equation}
  \begin{split}
    H(\textbf{X}) = \sum_{j=1}^n-p_jlog_2p_j
  \end{split}
 \end{equation}
Also, let $\{0,1\}^*$ be finite binary codewords of arbitrary length and $g: \textbf{X} \rightarrow \{0,1\}^*$ then average bit length of codeword is given as:
\begin{equation}
  \begin{split}
    L(g) = \sum_{j=1}^np_jl_j
  \end{split}
 \end{equation}
 where $l_j$ is the length of $g(x_j)$. That is, function $g$ associates a binary codeword of some length $l_j$ to $x_j$.
 \section{EXPERIMENTAL SETUP AND RESULTS }
\subsection{Dataset}
\begin{figure*}[!htbp]
  \centering
  \includegraphics[width=1.0\linewidth]{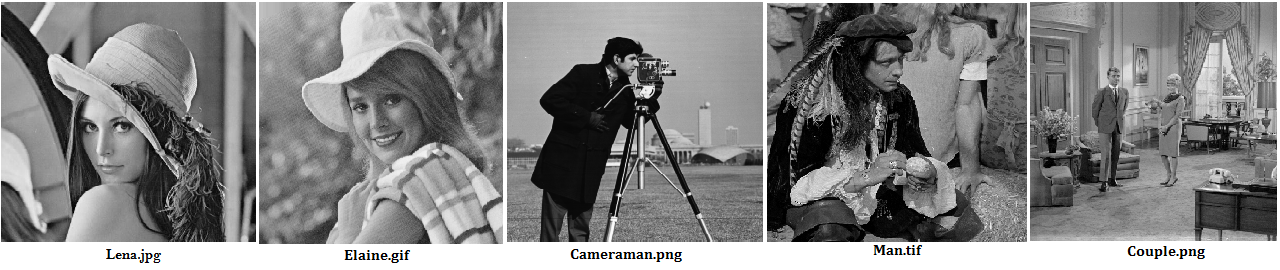}
  \captionsetup{justification=centering}
  \caption{Image Dataset 1}\label{example_composite}
\end{figure*}

\begin{figure*}[!htbp]
  \centering
  \includegraphics[width=1.0\linewidth]{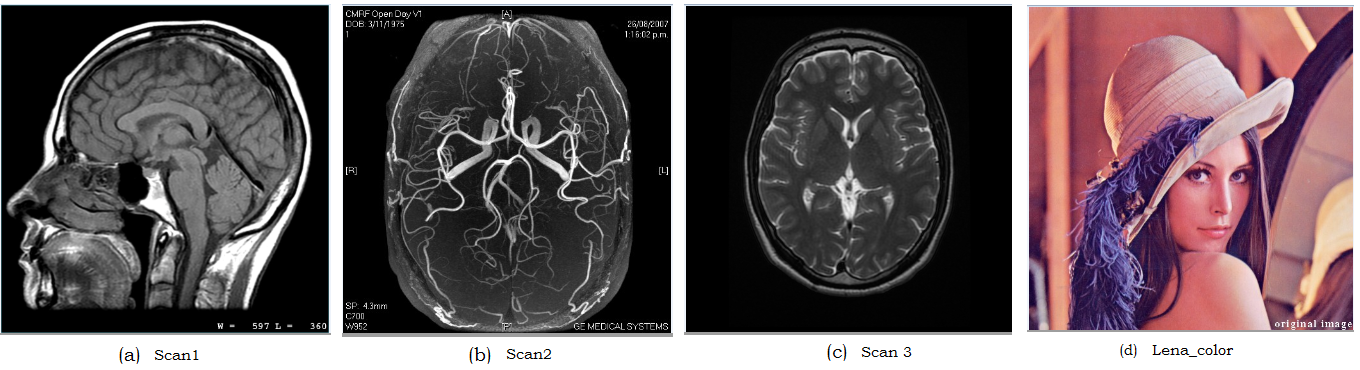}
  \captionsetup{justification=centering}
  \caption{Image Dataset 2}\label{example_3d}
\end{figure*}

\begin{figure*}[!htbp]
  \centering
  \includegraphics[width=1.0\linewidth]{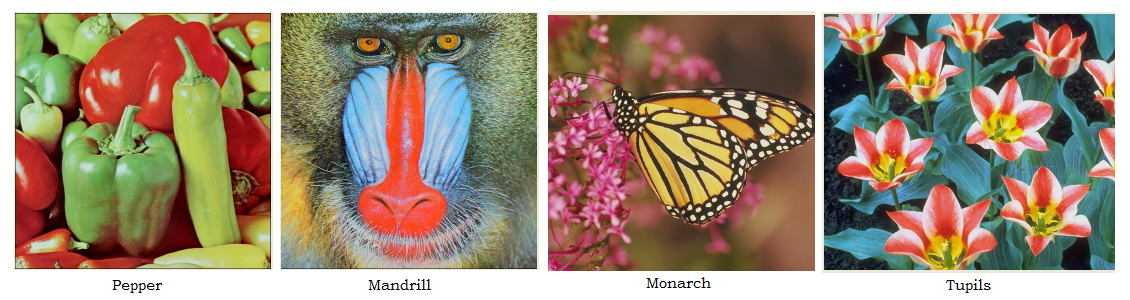}
  \captionsetup{justification=centering}
  \caption{Image Dataset 3}\label{new_img}
\end{figure*}

In the first set of experiments, five benchmark grayscale images in Fig~\ref{example_composite} were used. The images have dimensions: \emph{lena.jpg- size 512$\times$512}, \emph{elaine.gif- size 512$\times$512}, \emph{cameraman.png- size 256$\times$256}, \emph{man.tif- size 512$\times$512}, and \emph{couple.png- size 512$\times$512}.
It is generally remarked in medical imaging that 3D images can be viewed as time sequence of radiographic images or volume of tomographic slice image of a static object or tomographic slice images of a dynamic object \cite{sriraam20113,herman20003d}. The 3-D images considered in this work can be visualized as a stack of 2D image slices of a progressive variation of static object. In a way, 3D medical image data is abstracted as a 3D rectangular block of voxels, where each voxel has its assigned value \cite{sriraam20113}. The 3-D medical image data samples considered in this paper are: scan 1 \cite{scan1}, scan 2 is a  3D MRI image of the circle of willis and other cerebral arteries \cite{scan2}, and scan 3 is the axial T2-weighted MR image of a normal brain at the level of the lateral ventricles. These images are shown in Fig.~\ref{example_3d}a, b, and c respectively. In the second set of experiments, dataset 2 in Fig.~\ref{example_3d} were utilized. In the last set of experiments, dataset 3 in Fig.~\ref{new_img} which comprises of popular benchmark images are used to show the effectiveness of the proposed heuristics.

\begin{figure*}[htb!]
	\begin{minipage}[b]{0.24\linewidth}
		\centering
		\centerline{\includegraphics[scale=0.41]{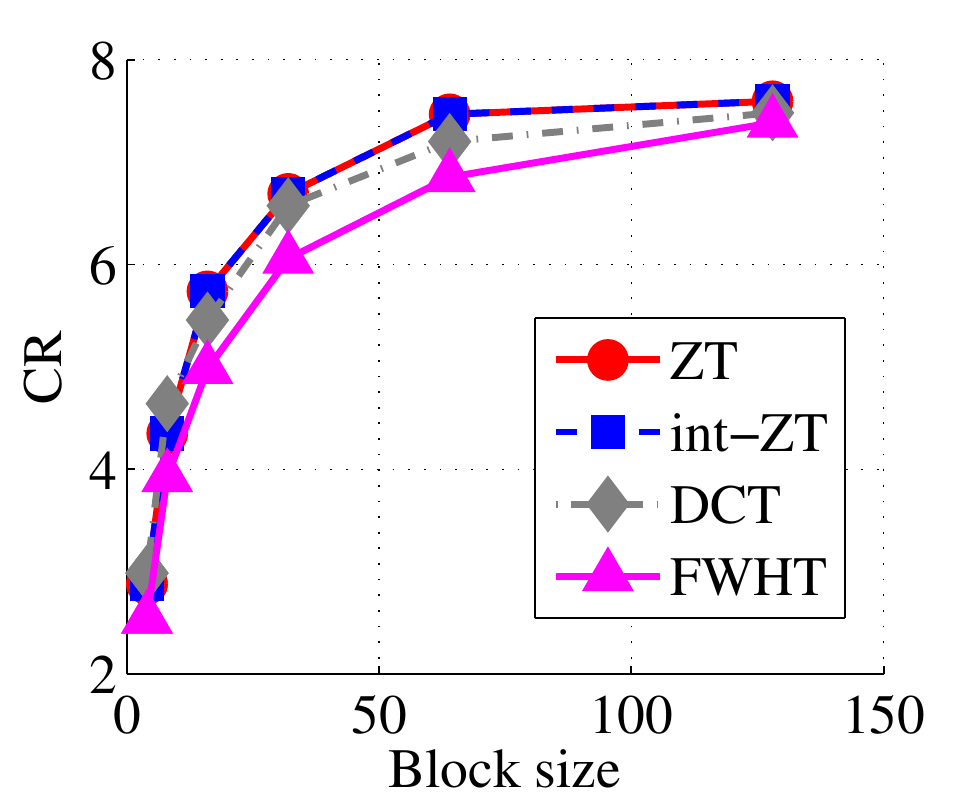}}
		{{\footnotesize (a)}}
	\end{minipage}
\begin{minipage}[b]{0.24\linewidth}
		\centering
		\centerline{\includegraphics[scale=0.41]{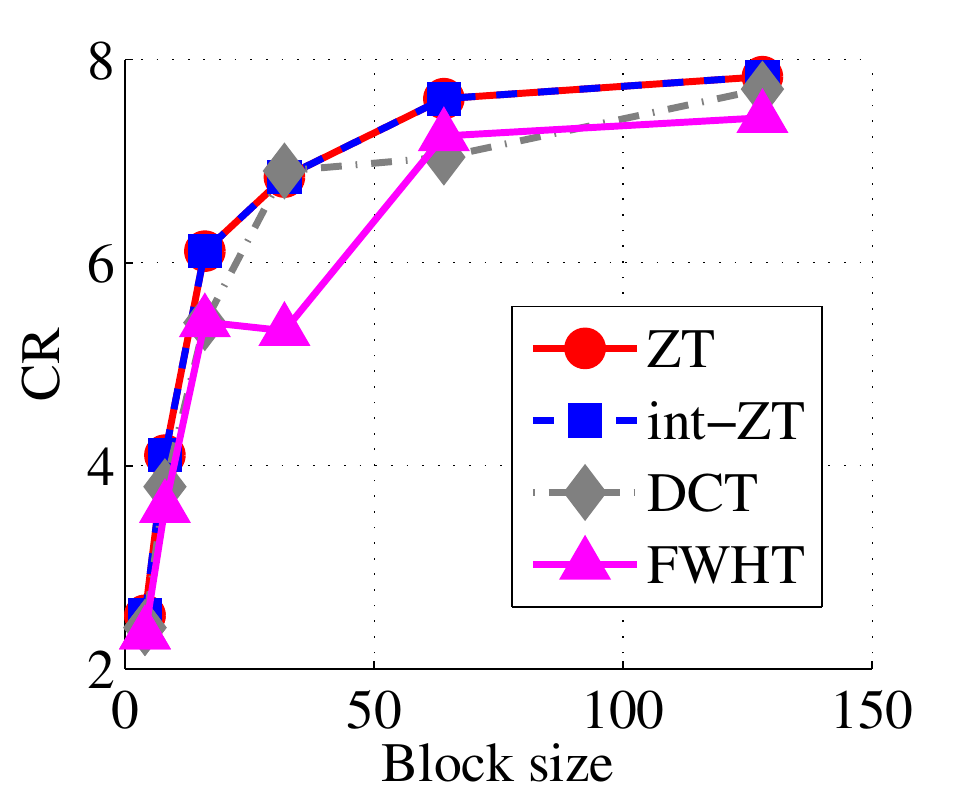}} %
		{{\footnotesize(b)}}
	\end{minipage}
\begin{minipage}[b]{0.24\linewidth}
		\centering
		\centerline{\includegraphics[scale=0.42]{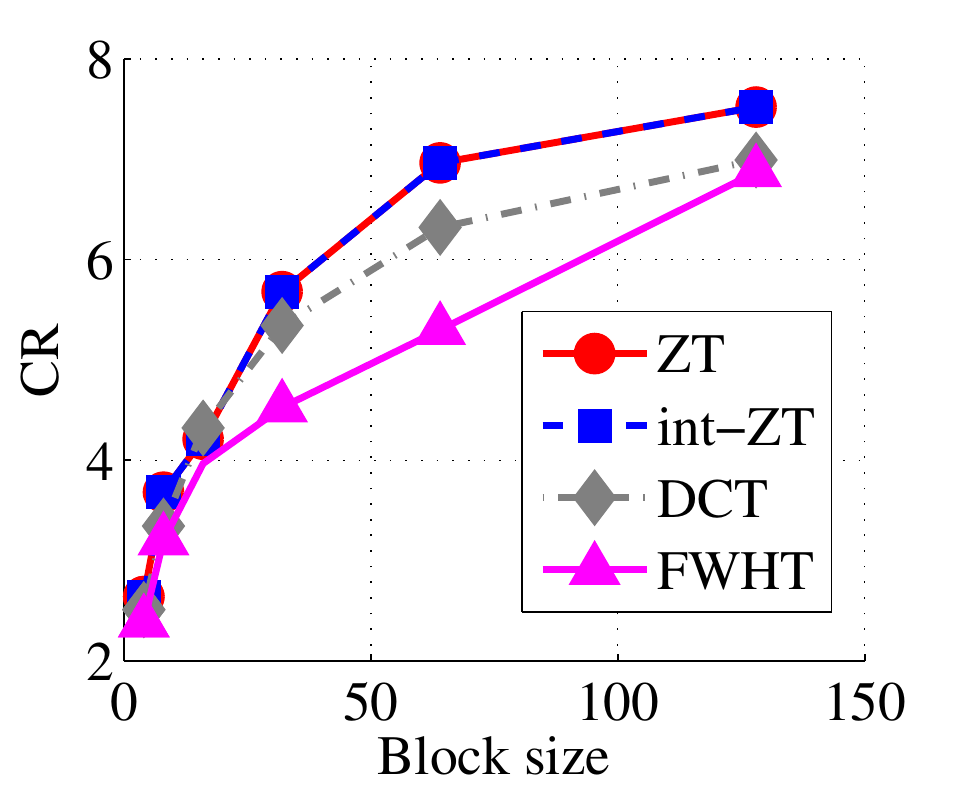}}
		{{\footnotesize (c)}}
	\end{minipage}
\begin{minipage}[b]{0.24\linewidth}
		\centering
		\centerline{\includegraphics[scale=0.42]{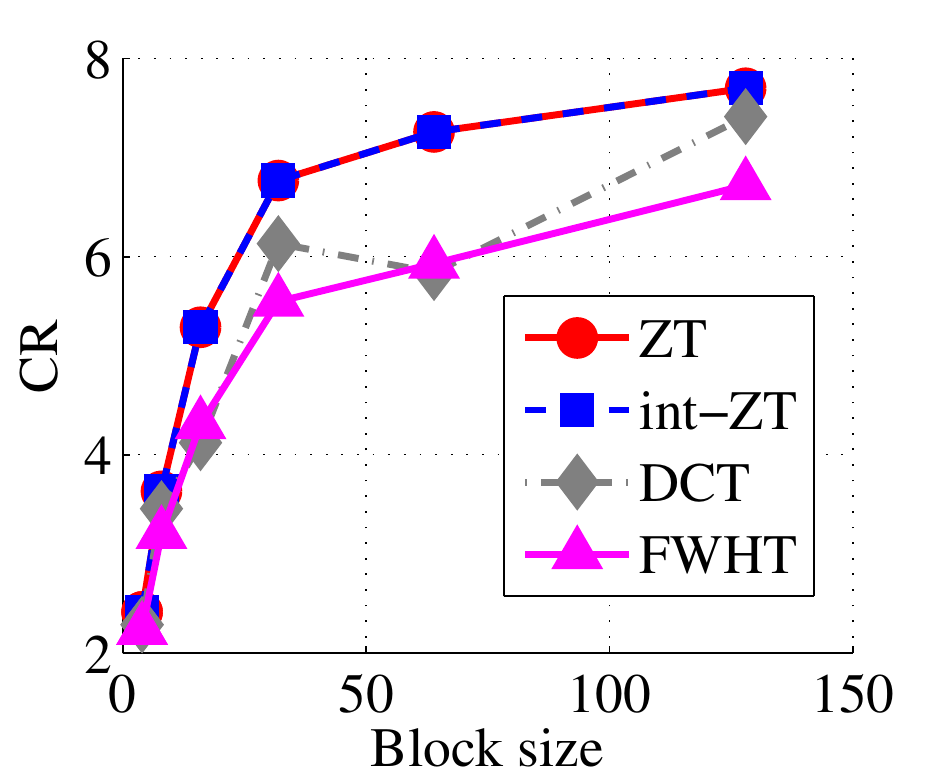}} %
		{{\footnotesize(d)}}
	\end{minipage}
\captionsetup{justification=centering}
	\caption{A plot of compression ratio against block size for DCT, FWHT, Zipper, and interlace-zipper transform using (a) Lena (b) Elaine (c) Cameraman (d) Couple from Dataset 1}\label{cr_results_dataset1}
\end{figure*}

\begin{figure*}[htb!]
	\begin{minipage}[b]{0.24\linewidth}
		\centering
		\centerline{\includegraphics[scale=0.41]{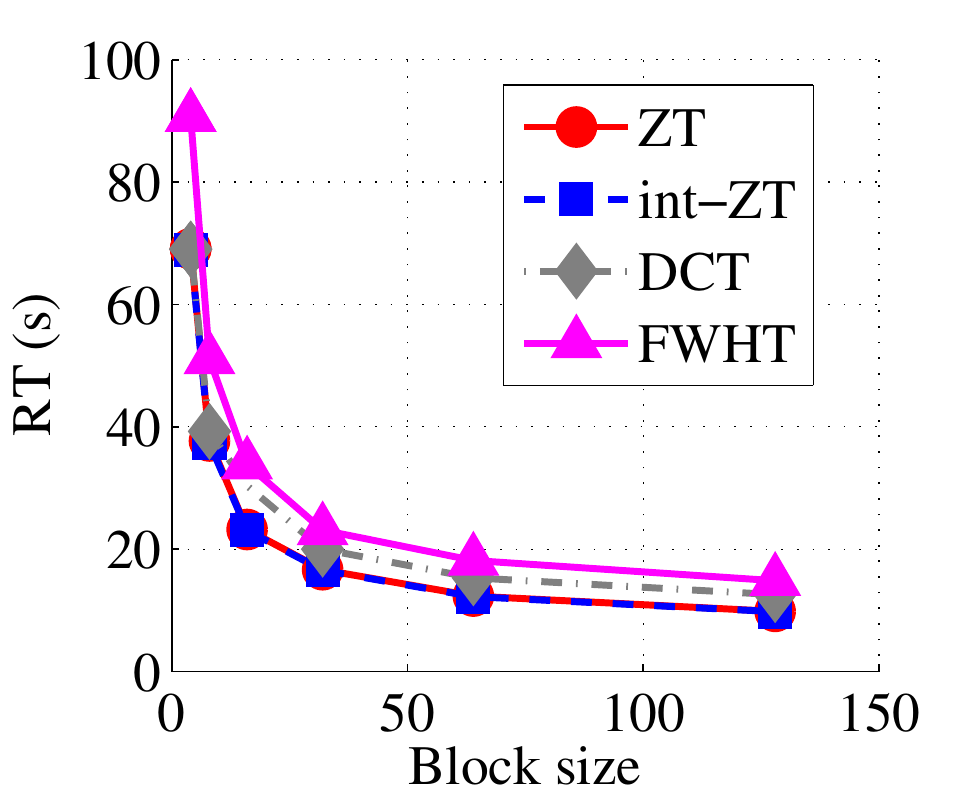}}
		{{\footnotesize (a)}}
	\end{minipage}
\begin{minipage}[b]{0.24\linewidth}
		\centering
		\centerline{\includegraphics[scale=0.41]{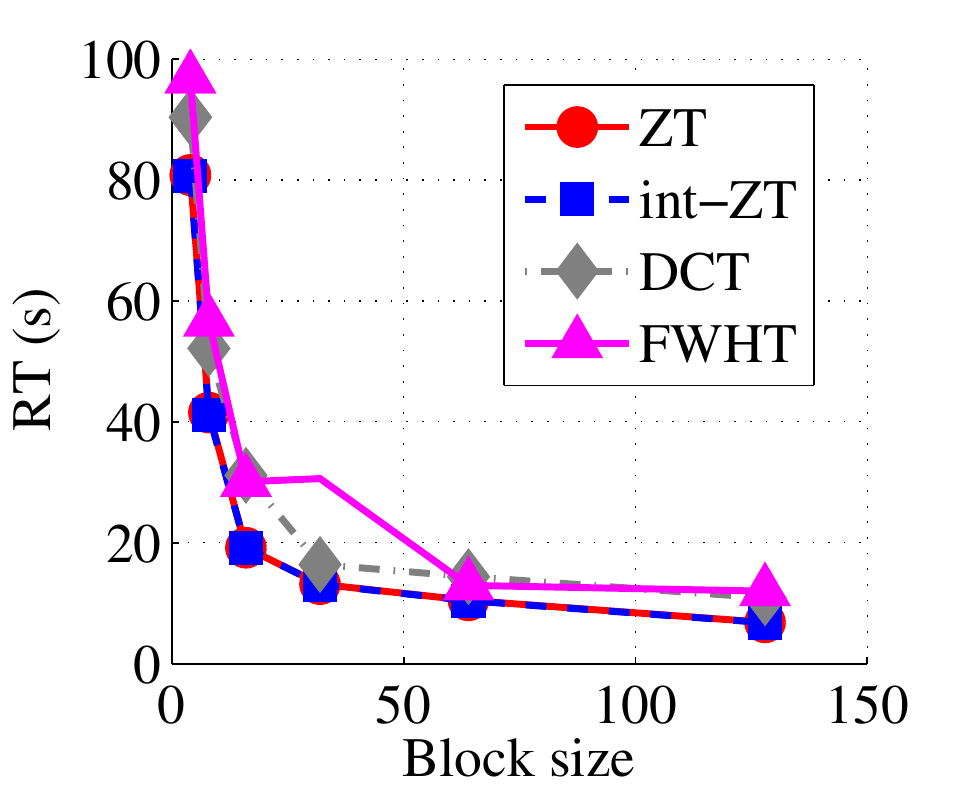}} %
		{{\footnotesize(b)}}
	\end{minipage}
\begin{minipage}[b]{0.24\linewidth}
		\centering
		\centerline{\includegraphics[scale=0.42]{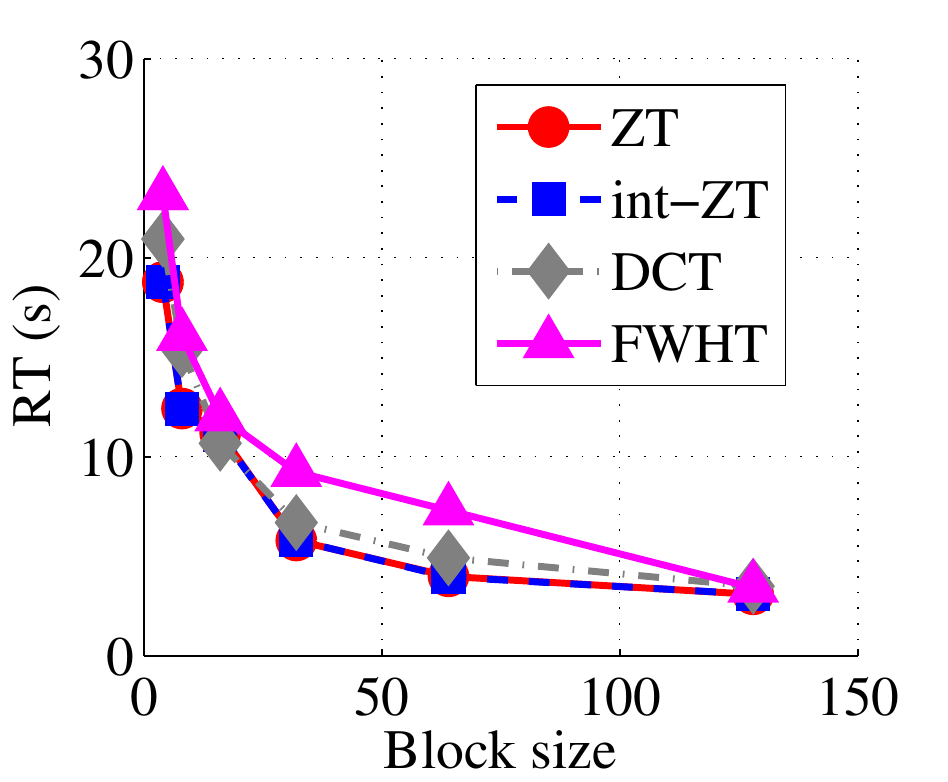}}
		{{\footnotesize (c)}}
	\end{minipage}
\begin{minipage}[b]{0.24\linewidth}
		\centering
		\centerline{\includegraphics[scale=0.42]{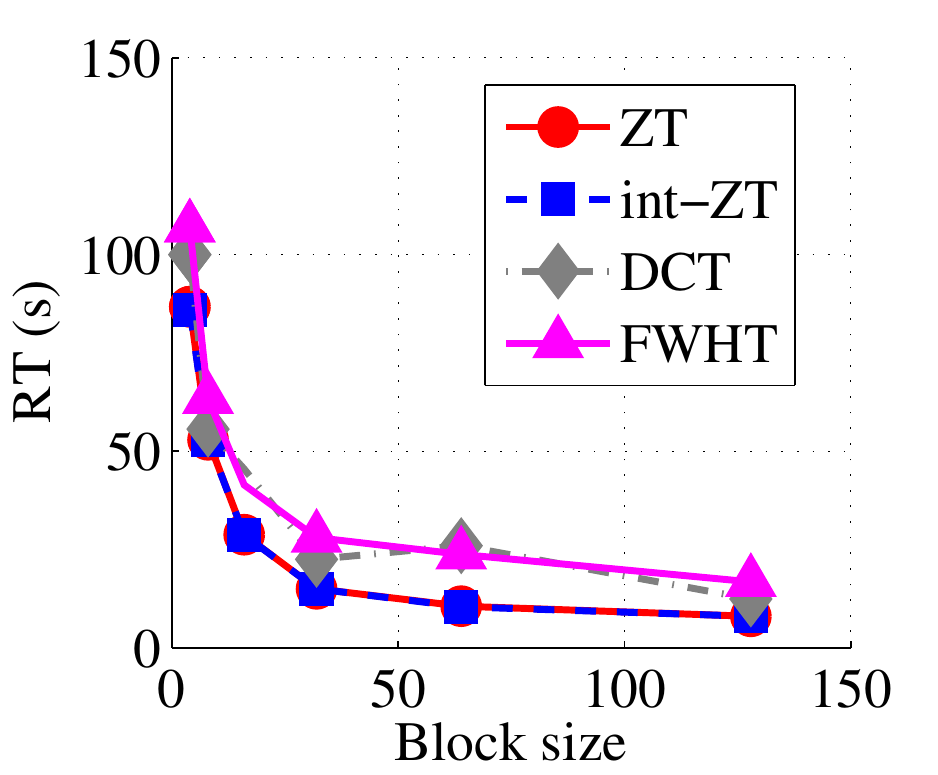}} %
		{{\footnotesize(d)}}
	\end{minipage}
\captionsetup{justification=centering}
	\caption{A plot of running time against block size for DCT, FWHT, Zipper, and interlace-zipper transform using (a) Lena (b) Elaine (c) Cameraman (d) Couple from Dataset 1}\label{rt_results_dataset1}
\end{figure*}
\subsection{Experimental Design}
The experiments were performed in MATLAB environment and we report the standard metrics: the compression ratio ($CR$) and the running time in (seconds) evaluated on a machine with Intel(r) Core(TM) i7-6700 CPU @ 3.40Ghz and a 64GB of RAM running a 64-bit Windows 10 Enterprise edition. The compression ratio ($CR$) is an important criterion in choosing a compression scheme. This criterion is used to compare different compression paradigms, and is defined as:
\begin{equation} \label{MyEq1}
  CR = \frac{\text{Original file size}}{\text{Compressed file size}}
\end{equation}
In order to benchmark the proposed scheme with other methods, we also implemented DCT and FWHT-based compression algorithms. We also utilized the running time to compare the proposed method with those of DCT and FWHT. In this work, we define the running time as the time elapsed between the zipper transformation and the inverse transformation including Huffman coding and decoding as shown in Fig.~\ref{zipper}. The MATLAB implementation of these algorithms can be downloaded from \url{https://github.com/babajide07/Zipper-Transformation}.\\
\\
\textbf{Experiment 1}. The objective of this experiment was to demonstrate the effectiveness of zipper transformation in the proposed compression scheme on grayscale images and also to demonstrate the exactness in performance of both concatenating-zipper and interlacing-zipper transforms. The data sets used were \emph{lena.jpg}, \emph{elaine.gif}, \emph{cameraman.png}, \emph{man.tif}, and \emph{couple.png} as shown in Fig.~\ref{example_composite}.\\
\\
\textbf{Experiment 2}. This experiment aimed to compare the proposed compression scheme with two existing schemes using multi-dimensional images. Dataset 2 in Fig.~\ref{example_3d} and dataset 3 in Fig.~\ref{new_img} were used to benchmark the proposed with other schemes.
\subsection{Experimental Results and Discussion}
\begin{figure*}[htb!]
	\begin{minipage}[b]{0.24\linewidth}
		\centering
		\centerline{\includegraphics[scale=0.42]{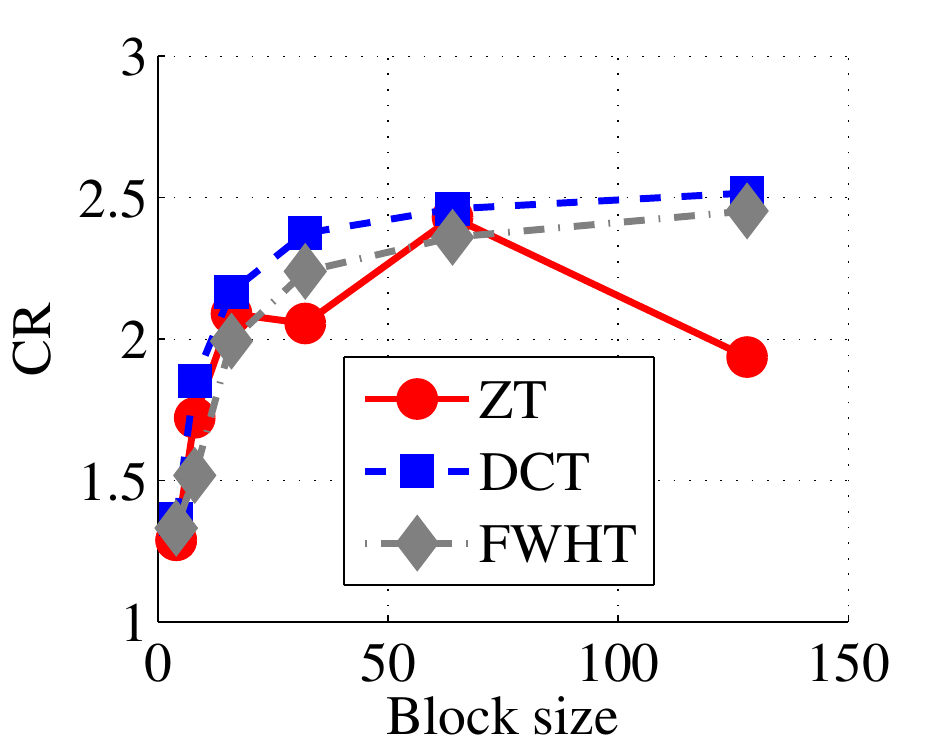}}
		{{\footnotesize (a)}}
	\end{minipage}
\begin{minipage}[b]{0.24\linewidth}
		\centering
		\centerline{\includegraphics[scale=0.42]{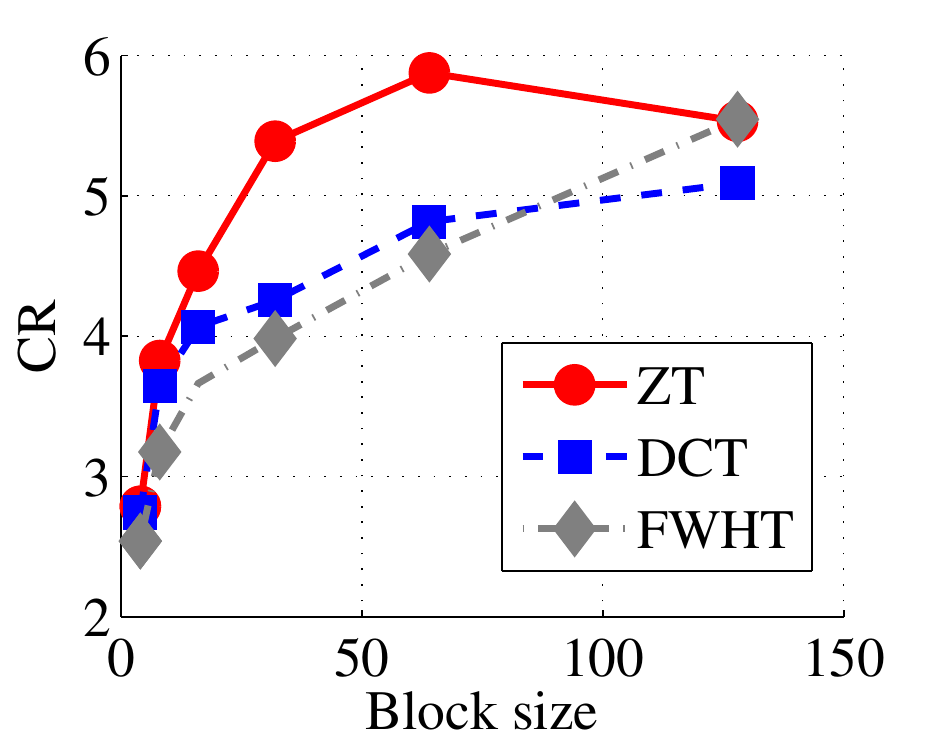}} %
		{{\footnotesize(b)}}
	\end{minipage}
\begin{minipage}[b]{0.24\linewidth}
		\centering
		\centerline{\includegraphics[scale=0.42]{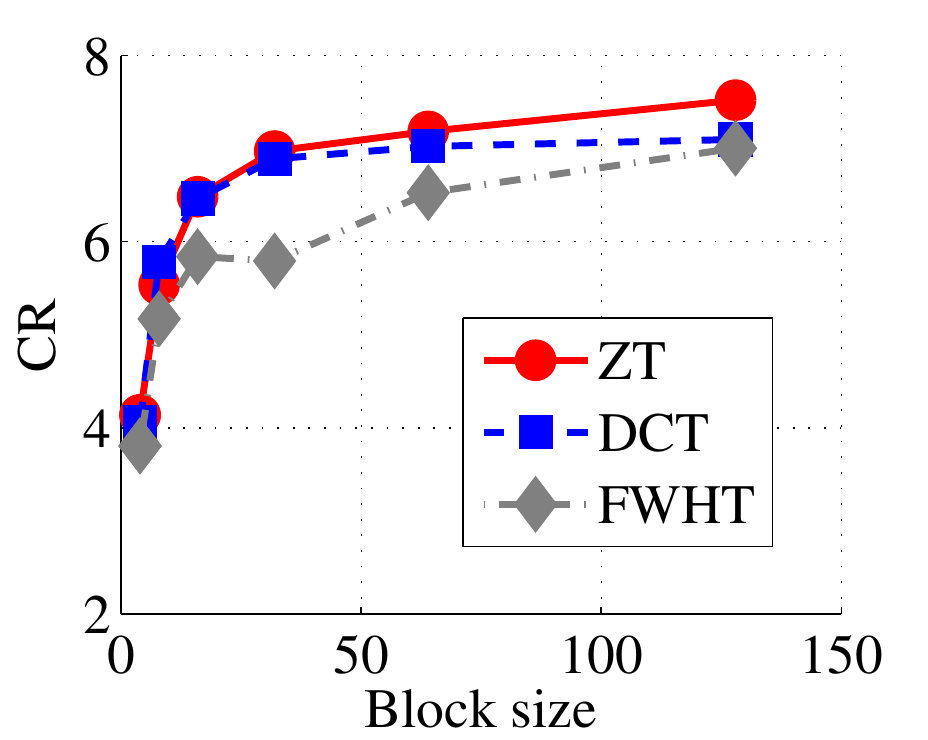}}
		{{\footnotesize (c)}}
	\end{minipage}
\begin{minipage}[b]{0.24\linewidth}
		\centering
		\centerline{\includegraphics[scale=0.42]{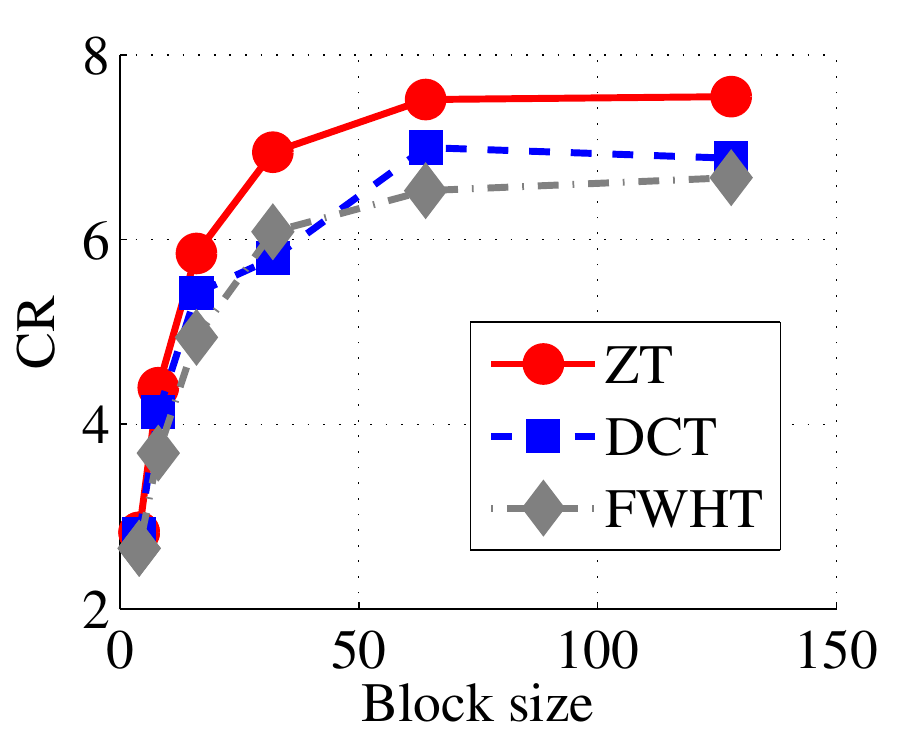}} %
		{{\footnotesize(d)}}
\end{minipage}
\captionsetup{justification=centering}
	\caption{A plot of compression ratio against block size for DCT, FWHT, Zipper, and interlacing-zipper transform using (a) Scan1 (b) Scan2 (c) Scan3 (d) Lena\_color from Dataset 2}\label{cr_results_dataset2}
\end{figure*}

\begin{figure*}[htb!]
	\begin{minipage}[b]{0.24\linewidth}
		\centering
		\centerline{\includegraphics[scale=0.42]{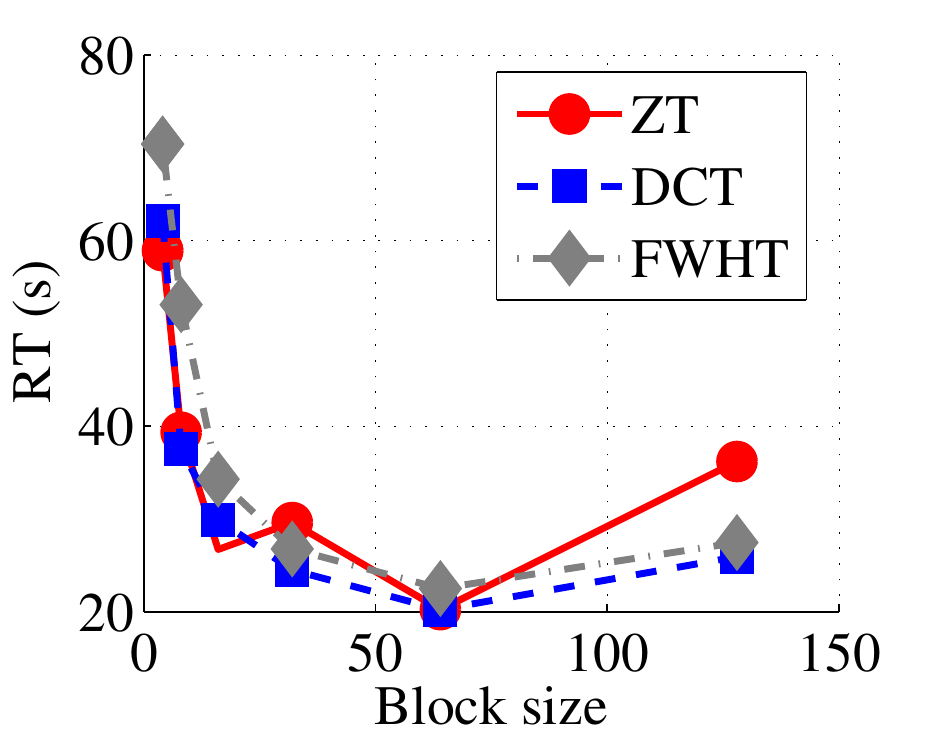}}
		{{\footnotesize (a)}}
	\end{minipage}
\begin{minipage}[b]{0.24\linewidth}
		\centering
		\centerline{\includegraphics[scale=0.42]{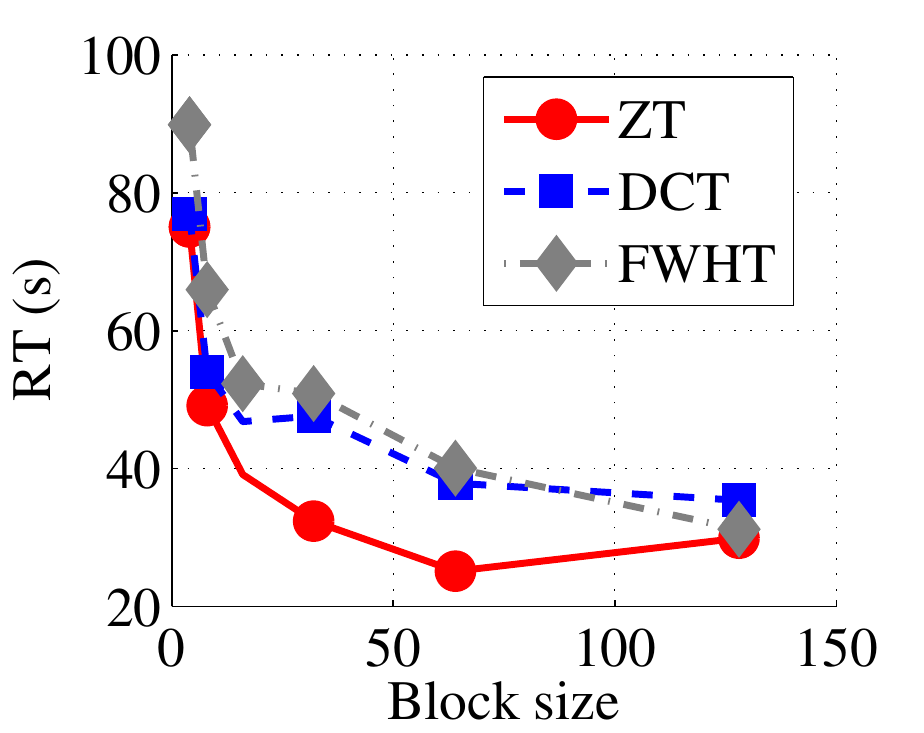}} %
		{{\footnotesize(b)}}
	\end{minipage}
\begin{minipage}[b]{0.24\linewidth}
		\centering
		\centerline{\includegraphics[scale=0.42]{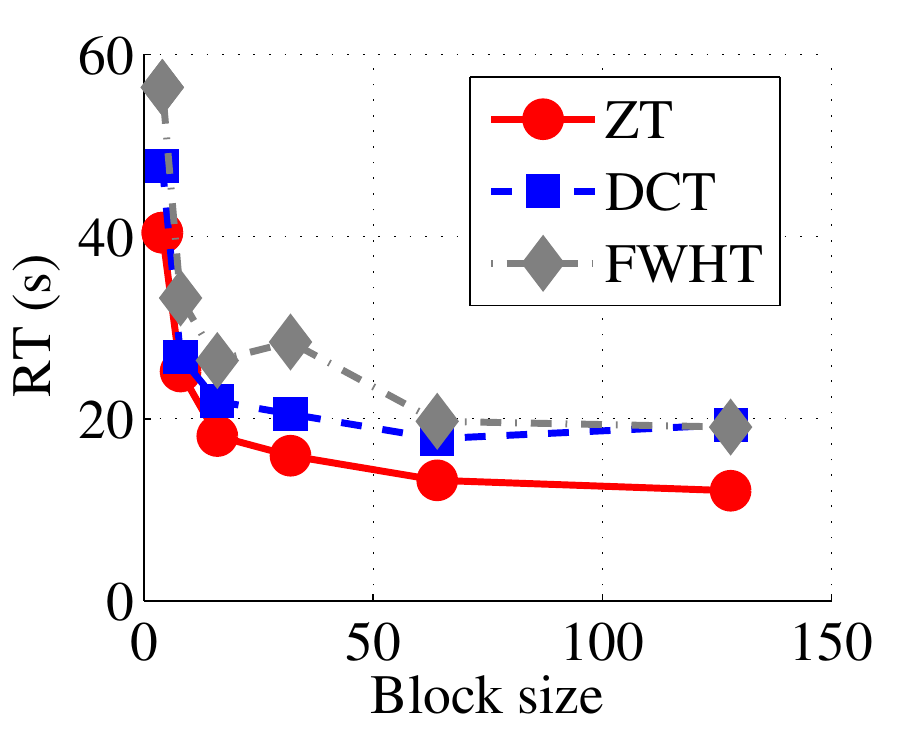}}
		{{\footnotesize (c)}}
	\end{minipage}
\begin{minipage}[b]{0.24\linewidth}
		\centering
		\centerline{\includegraphics[scale=0.42]{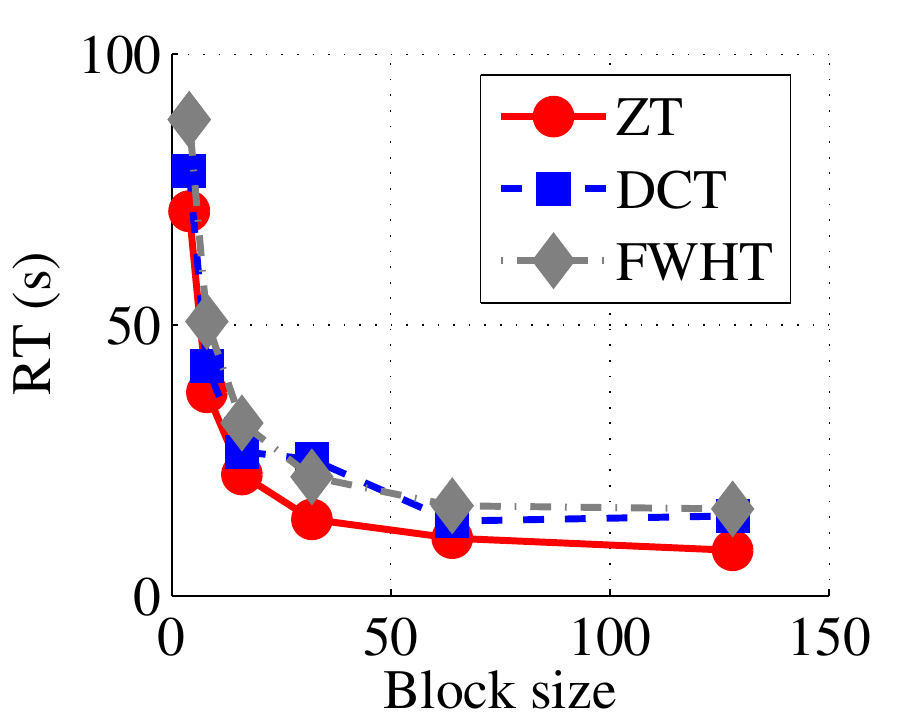}} %
		{{\footnotesize(d)}}
	\end{minipage}
\captionsetup{justification=centering}
	\caption{A plot of running time against block size for DCT, FWHT, Zipper, and interlacing-zipper transform using (a) Scan1 (b) Scan2 (c) Scan3 (d) Lena\_color from Dataset 2}\label{rt_results_dataset2}
\end{figure*}

\begin{figure*}[htb!]
	\begin{minipage}[b]{0.24\linewidth}
		\centering
		\centerline{\includegraphics[scale=0.42]{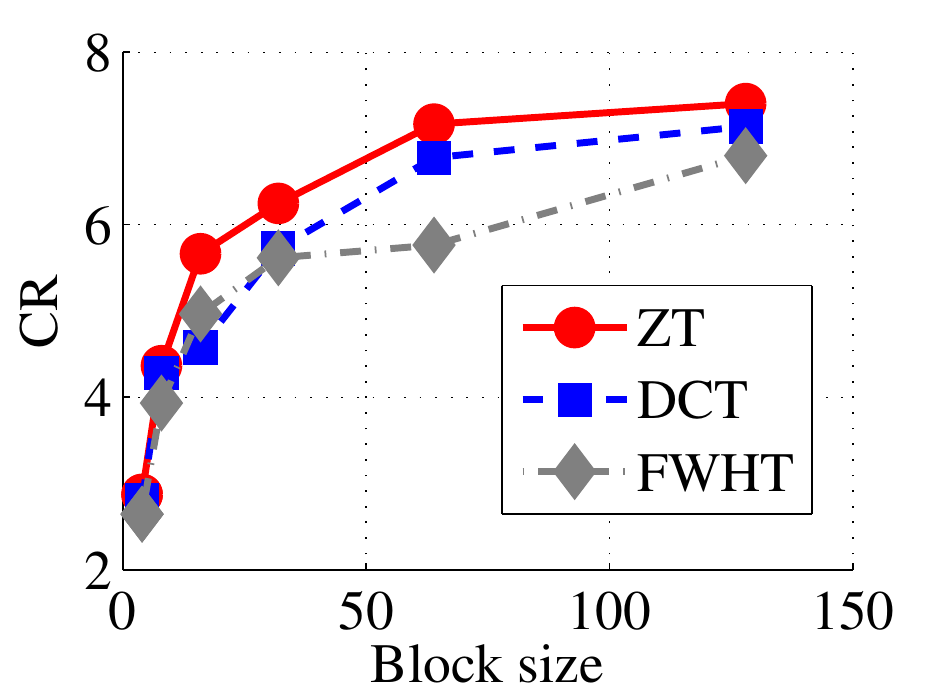}}
		{{\footnotesize (a)}}
	\end{minipage}
\begin{minipage}[b]{0.24\linewidth}
		\centering
		\centerline{\includegraphics[scale=0.42]{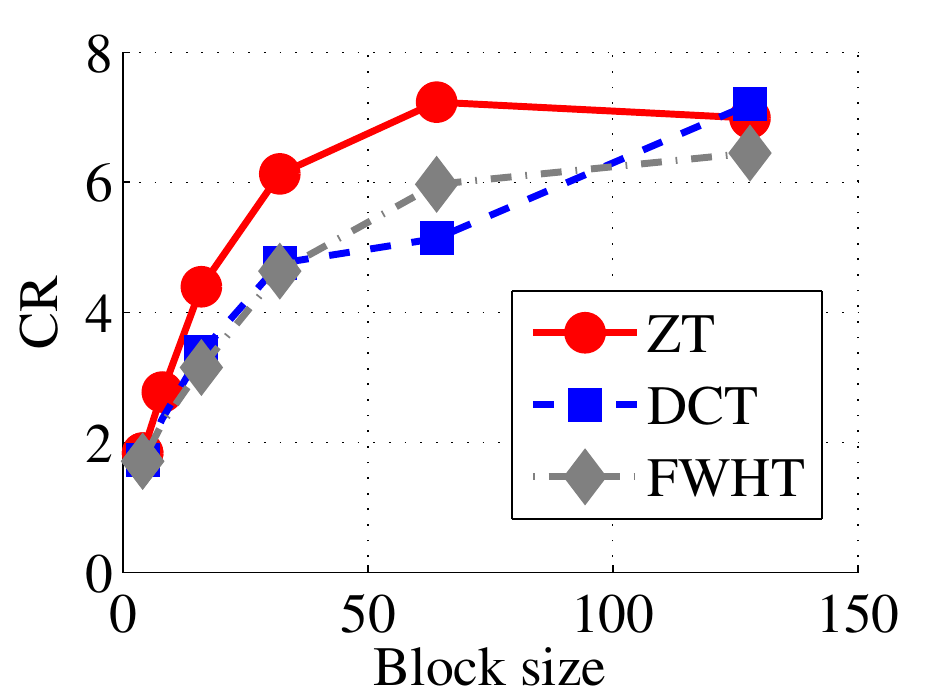}} %
		{{\footnotesize(b)}}
	\end{minipage}
\begin{minipage}[b]{0.24\linewidth}
		\centering
		\centerline{\includegraphics[scale=0.42]{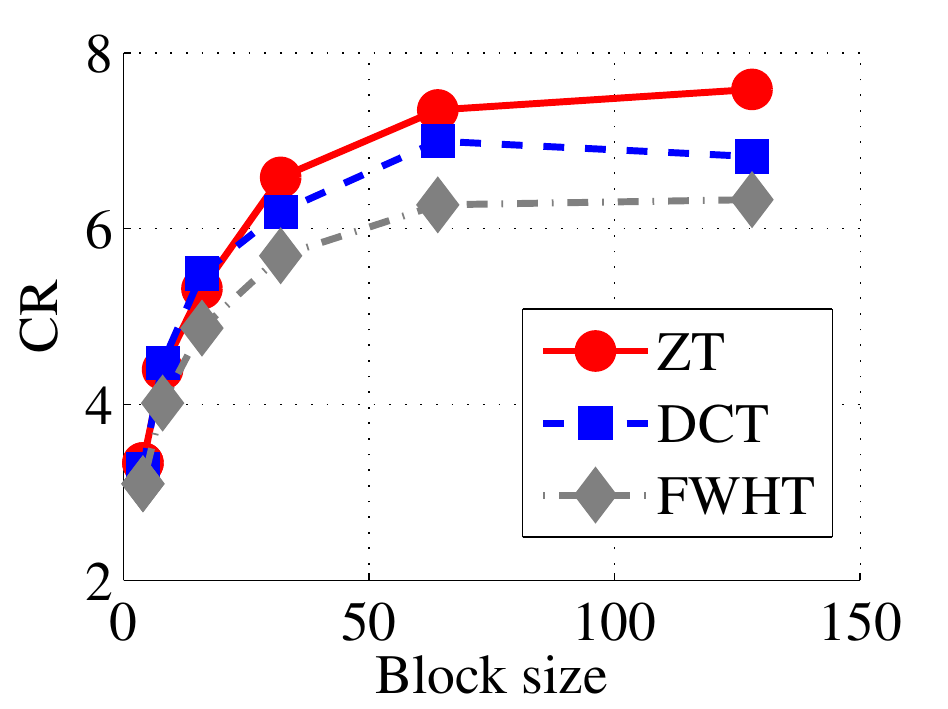}}
		{{\footnotesize (c)}}
	\end{minipage}
\begin{minipage}[b]{0.24\linewidth}
		\centering
		\centerline{\includegraphics[scale=0.42]{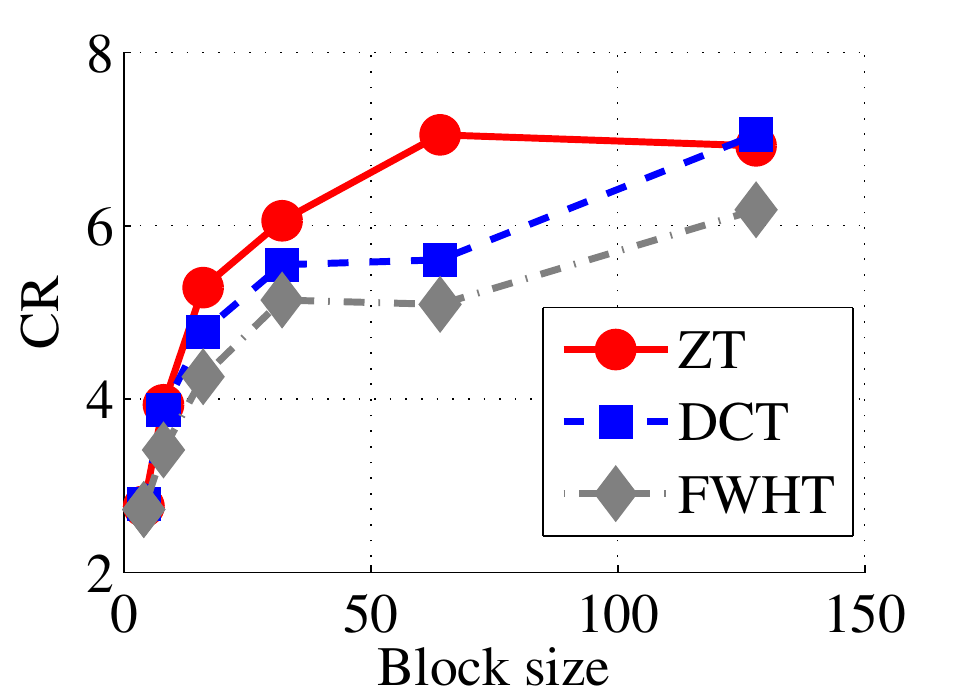}} %
		{{\footnotesize(d)}}
	\end{minipage}
\captionsetup{justification=centering}
	\caption{A plot of compression ratio against block size for DCT, FWHT, Zipper, and interlacing-zipper transform using (a) Pepper (b) Mandrill (c) Monarch (d) Tulips from Dataset 3}\label{cr_results_dataset3}
\end{figure*}

\begin{figure*}[htb!]
	\begin{minipage}[b]{0.24\linewidth}
		\centering
		\centerline{\includegraphics[scale=0.42]{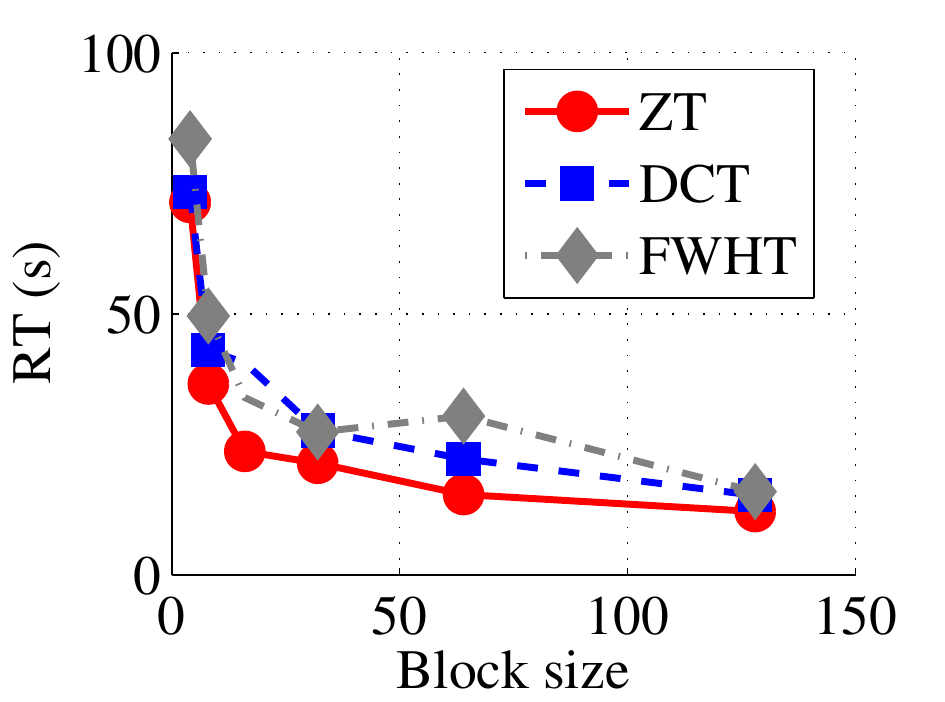}}
		{{\footnotesize (a)}}
	\end{minipage}
\begin{minipage}[b]{0.24\linewidth}
		\centering
		\centerline{\includegraphics[scale=0.42]{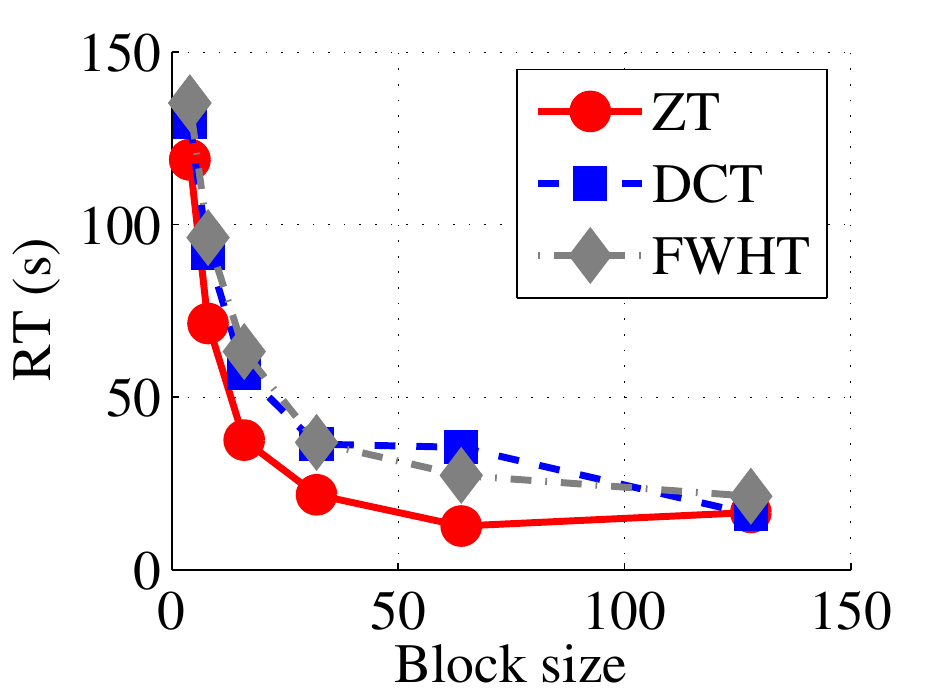}} %
		{{\footnotesize(b)}}
	\end{minipage}
\begin{minipage}[b]{0.24\linewidth}
		\centering
		\centerline{\includegraphics[scale=0.42]{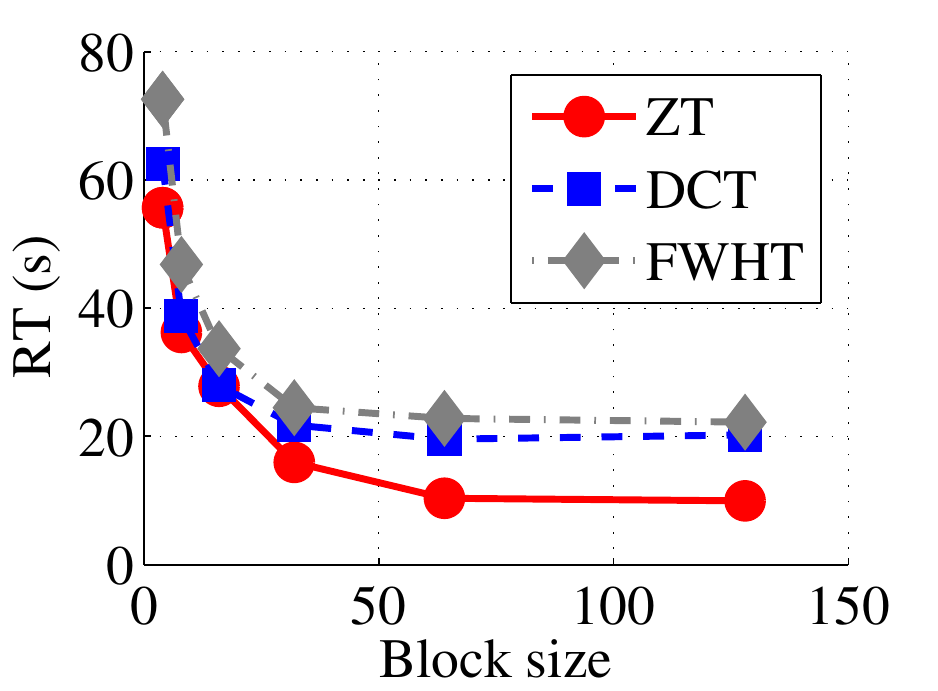}}
		{{\footnotesize (c)}}
	\end{minipage}
\begin{minipage}[b]{0.24\linewidth}
		\centering
		\centerline{\includegraphics[scale=0.42]{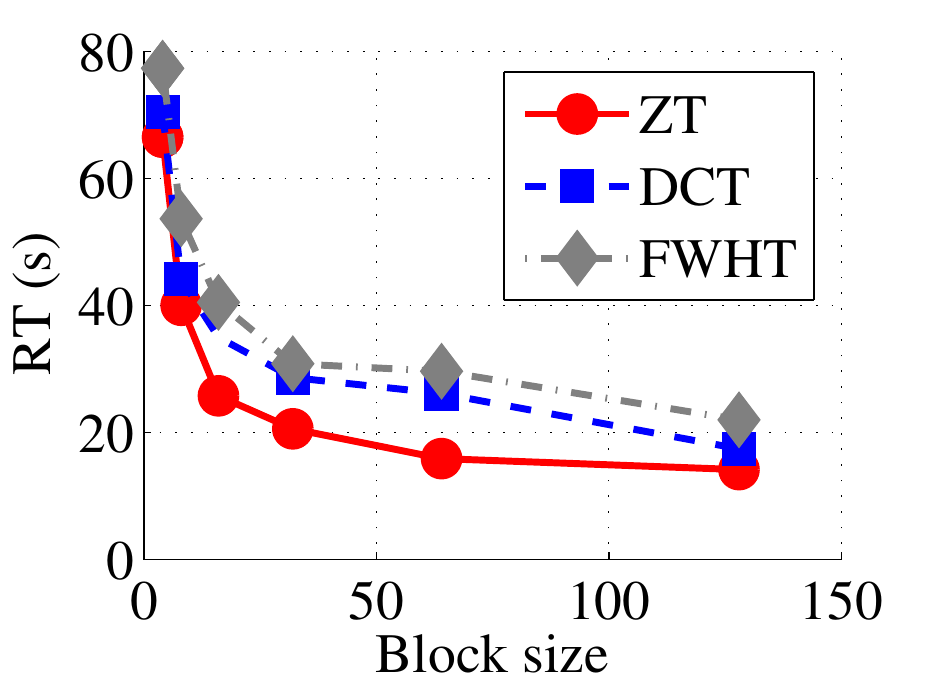}} %
		{{\footnotesize(d)}}
	\end{minipage}
\captionsetup{justification=centering}
	\caption{A plot of running time against block size for DCT, FWHT, Zipper, and interlacing-zipper transform using (a) Pepper (b) Mandrill (c) Monarch (d) Tulips from Dataset 3}\label{rt_results_dataset3}
\end{figure*}

\textbf{Experiment 1}: In the first set of experiments using Lena image, both the ZT-based and interlacing-ZT-based compression algorithms outperform both the DCT and FWHT counterparts in terms of CR for many block sizes as seen in Fig.~\ref{cr_results_dataset1}a. On the average, both the ZT-based and interlacing-ZT-based compression algorithms significantly compress the lena.jpg image better than DCT and FWHT. In addition to efficient compression of Lena image, ZT and interlacing-ZT have faster implementations than DCT and FWHT as shown in Fig.~\ref{rt_results_dataset1}a. The proposed algorithms were also tested using the Elaine.gif image and the compression ratio and running time are depicted in Figs.~\ref{cr_results_dataset1}b and \ref{rt_results_dataset1}b respectively. Again, we observed that ZT and interlacing-ZT outperform their counterparts in both compressing capabilities and speed of implementation.\\
\indent
In the next set of experiments, we benchmark the proposed schemes with DCT and FWHT in terms of CR and running time using the cameraman.png image data. It was observed in Figs.~\ref{cr_results_dataset1}c and \ref{rt_results_dataset1}c that the performances of ZT and interlacing-ZT outweighs those of DCT and FWHT. For couple.png image data, we again observe that ZT and interlacing-ZT are superior to both DCT and FWHT in term of CR for most of the block sizes considered as shown in Figs.~\ref{cr_results_dataset1}d. In fact, a close scrutiny of the results shows that the performance is more pronounced when the block size is 64. In addition to the poor compressing capability, implementing DCT and FWHT are more expensive in terms of running time than the ZT-based compression paradigms as shown in Fig.~\ref{rt_results_dataset1}d.\\
\indent
In Tables~\ref{table:1} and \ref{table:2}, we report the average length of the codewords and average entropy for Dataset 1 in Fig.~\ref{example_composite} for all the four methods considered. Each image in Dataset 1 is transformed using DCT, FHWT, Zipper, and interlacing-Zipper then encoded using Huffman coding scheme. Entropy and average length of the codewords were recorded and averaged over all the images in Dataset 1. The most efficient of the four methods is highlighted in bold fonts. Our observations for all the experiments reported in Tables~\ref{table:1} and \ref{table:2} are that average length of the codeword decreases as the entropy decreases for all the four methods, and also the value of entropy is very close to the average codeword length. Closer look at each of the compression method reveals that the discrepancy between the average length of the codeword and the entropy is smallest for ZT-based algorithms compared to DCT and FWHT-based methods. In addition, the average length for zipper-based transforms is less than those of DCT and FWHT for almost all the block sizes tested. It is also remarked that both ZT and interlacing-ZT yield similar results which might indicate robustness of the proposed approach. \\
\\
\textbf{Experiment 2}: In this set of experiments, we evaluated the ability of the proposed method to compress high dimensional data. For these experiments, the proposed compression method was deployed on each dimension and the performance metrics are averaged over all the dimensions. In this regard, Dataset 2 ( comprising of three medical images and the color version of lena.jpg image) was used to benchmark ZT-based algorithm with those of DCT and FWHT. It can be observed in Figs.~\ref{cr_results_dataset2}a and \ref{rt_results_dataset2}a that ZT-based transform have competitive performance on Scan1 data for block sizes 4$\times$4, 16$\times$16, and 64$\times$64. However for 8$\times$8, 32$\times$32, and 128$\times$128, performance of ZT-based algorithm plummeted in comparison with DCT and FWHT-based approaches. Experimentation with Scan2 image shows that ZT consistently compresses the MRI data better and faster than its counterpart for most of the block sizes considered as shown in Figs.~\ref{cr_results_dataset2}b and \ref{rt_results_dataset2}b. With block size of 128$\times$128, the performance of FWHT improves and became comparable with ZT. Better performance in compression and speed of implementation is again observed on the average for ZT-based method using the Scan3 image, and the performance is more pronounced when the block size is 128$\times$128 as shown in Figs.~\ref{cr_results_dataset2}c and \ref{rt_results_dataset2}c. Again the proposed algorithm using color version of Lena image was compared with those of DCT and FWHT. It was observed that for most of the block sizes, ZT consistently compresses better and faster than its counterparts as shown in Figs.~\ref{cr_results_dataset2}d and \ref{rt_results_dataset2}d. For block size $4$ and $8$, zipper transform has slightly better compression capability and running time than DCT, however as the block size increases, performance of ZT became more pronounced.\\
\indent
In the last set of experiments, we evaluated the performance of the proposed algorithm on popular color image collection tagged in this paper as Dataset 3 and the performance is compared with those of DCT and FWHT. It is remarked that ZT-based algorithm outperforms its counterpart on the average in terms of compression ratio and algorithmic running time (in seconds) for all the four color images in this particular dataset as shown in Figs.~\ref{cr_results_dataset3} and \ref{rt_results_dataset3} respectively. Also, Tables~\ref{table:4} and \ref{table:3} respectively give the synopsis of the average length of the codewords and average entropy of the compressed version of the images in Dataset3 using ZT, DCT, and FWHT. Once again, the most efficient of the three methods is highlighted in bold fonts. There is strong correlation between the average codeword and average entropy for all the three heuristics as inferred in Tables~\ref{table:3} and \ref{table:4}. Again, the disparity between the average length of the codeword and the entropy is least in ZT-based algorithms in comparison with DCT and FWHT-based methods. Furthermore, the average length and entropy for zipper-based transforms is less than those of DCT and FWHT for almost all the block sizes tested.

\begin{table}[htb!]
 \setlength{\tabcolsep}{4pt}
	\caption{Average Entropy for DCT, FWHT and ZT using Image Dataset 1}
    \centering
	\scalebox{0.85}{
		\begin{tabular}{|c|c c|c c |c c|c c|}
			\hline
			\multicolumn{1}{|c|}{} & \multicolumn{2}{c|}{ DCT} & \multicolumn{2}{c|}{FWHT} & \multicolumn{2}{|c|}{Zipper} & \multicolumn{2}{|c|}{interlacing-Zipper}\\
			\hline
			\multicolumn{1}{|c|}{\textit{Block Sizes}} & \textit{Average} & ($\pm$ \textit{STD}) & \textit{Average} & ($\pm$ \textit{STD})& \textit{Average} & ($\pm$ \textit{STD}) & \textit{Average} & ($\pm$ \textit{STD})\\
			\hline
			4 $\times$ 4 & 3.3407 & 0.3811 &  3.4225 & 0.3314 & \textbf{3.1884} &  0.3277 & \textbf{3.1884} &  0.3277\\
            8 $\times$ 8 & 2.1946 & 0.5071 &  2.4585 &  0.3422 & \textbf{1.9828} & 0.2578 & \textbf{1.9828} & 0.2578 \\
			16 $\times$ 16 & 1.4997 &  0.4684 & 1.8488 & 0.2704 & \textbf{1.3009} & 0.2231 & \textbf{1.3009} & 0.2231\\
			32 $\times$ 32 &  0.9383 & 0.2974 &  1.3007 &  0.2673 & \textbf{0.9223} &  0.3350 & \textbf{0.9223} & 0.3350\\
            64 $\times$ 64 & \textbf{0.5940} & 0.2396 &  0.9574 &  0.2141 & 0.6256 &  0.2672 & 0.6256 & 0.2672\\
            128 $\times$ 128 & 0.3754 &  0.1546 & 0.5368 & 0.2288 &  \textbf{0.3128} & 0.2084 & \textbf{0.3128} & 0.2084\\
			\hline
		\end{tabular}
\label{table:1}
}
\end{table}

\begin{table}[htb!]
 \setlength{\tabcolsep}{4pt}
	\caption{Average Length for DCT, FWHT and ZT using Image Dataset 1}
    \centering
	\scalebox{0.85}{
		\begin{tabular}{|c|c c|c c |c c|c c|}
			\hline
			\multicolumn{1}{|c|}{} & \multicolumn{2}{c|}{ DCT} & \multicolumn{2}{c|}{FWHT} & \multicolumn{2}{|c|}{Zipper} & \multicolumn{2}{|c|}{interlacing-Zipper}\\
			\hline
			\multicolumn{1}{|c|}{\textit{Block Sizes}} & \textit{Average} & ($\pm$ \textit{STD}) & \textit{Average} & ($\pm$ \textit{STD})& \textit{Average} & ($\pm$ \textit{STD}) & \textit{Average} & ($\pm$ \textit{STD})\\
			\hline
			4 $\times$ 4 &  3.3014 & 0.4208 & 3.4803 & 0.2819 & \textbf{3.1965} & 0.3412 & \textbf{3.1965} & 0.3412\\
            8 $\times$ 8 & 2.2606 & 0.3820 & 2.4522 & 0.3600 & \textbf{2.1028} & 0.2059 & \textbf{2.1028} & 0.2059\\
			16 $\times$ 16 & 1.7575 & 0.2698 &  1.8135 & 0.2682 &  \textbf{1.5459} & 0.2293 & \textbf{1.5459} & 0.2293\\
			32 $\times$ 32 & 1.3167 & 0.1372 & 1.5324 & 0.1723 & \textbf{1.2452} & 0.0992 & \textbf{1.2452} & 0.0992\\
            64 $\times$ 64 & 1.2173 & 0.1047 & 1.2785 & 0.1594 & \textbf{1.0980} & 0.0383 & \textbf{1.0980} & 0.0383 \\
            128 $\times$ 128 & 1.0896 &  0.0417 & 1.1423 & 0.0579 &  \textbf{1.0743} & 0.0685 & \textbf{1.0743} & 0.0685\\
			\hline
		\end{tabular}
\label{table:2}
}
\end{table}

\begin{table}[htb!]
 \setlength{\tabcolsep}{4pt}
	\caption{Average Entropy for DCT, FWHT and ZT using Image Dataset 3}
    \centering
	\scalebox{0.95}{
		\begin{tabular}{|c|c c|c c |c c|}
			\hline
			\multicolumn{1}{|c|}{} & \multicolumn{2}{c|}{ DCT} & \multicolumn{2}{c|}{FWHT} & \multicolumn{2}{|c|}{Zipper} \\
			\hline
			\multicolumn{1}{|c|}{\textit{Block Sizes}} & \textit{Average} & ($\pm$ \textit{STD}) & \textit{Average} & ($\pm$ \textit{STD})& \textit{Average} & ($\pm$ \textit{STD})\\
			\hline
			4 $\times$ 4 &  3.1406 & 1.0254 & 3.2781  & 0.9411 & \textbf{3.0653} & 0.8818  \\
            8 $\times$ 8 & 2.2848 & 0.7250 & 2.4803  & 0.6613 & \textbf{2.2116} & 0.4427 \\
			16 $\times$ 16 & 1.5497 &  0.5711  & 1.8245 & 0.4458 & \textbf{1.2538} & 0.4305 \\
			32 $\times$ 32 & 1.1548  & 0.3879 &  1.3593  & 0.3471 & \textbf{0.7525} & 0.1803  \\
            64 $\times$ 64 & 0.9357 & 0.1572 &  0.8544  & 0.2310 &  \textbf{0.6638} & 0.3021 \\
            128 $\times$ 128 & 0.6650 &  0.1194 &  0.6703 & 0.1776 &  \textbf{0.3816} & 0.0845 \\
			\hline
		\end{tabular}
\label{table:3}
}
\end{table}

\begin{table}[htb!]
 \setlength{\tabcolsep}{4pt}
	\caption{Average Length for DCT, FWHT and ZT using Image Dataset 3}
    \centering
	\scalebox{0.95}{
		\begin{tabular}{|c|c c|c c |c c|}
			\hline
			\multicolumn{1}{|c|}{} & \multicolumn{2}{c|}{ DCT} & \multicolumn{2}{c|}{FWHT} & \multicolumn{2}{|c|}{Zipper}\\
			\hline
			\multicolumn{1}{|c|}{\textit{Block Sizes}} & \textit{Average} & ($\pm$ \textit{STD}) & \textit{Average} & ($\pm$ \textit{STD})& \textit{Average} & ($\pm$ \textit{STD}) \\
			\hline
			4 $\times$ 4 & 3.2027 & 0.9767 & 3.3024 & 0.9318 & \textbf{3.1090} & 0.8575 \\
            8 $\times$ 8 & 2.2759 & 0.7473 & 2.4643 & 0.6990 & \textbf{2.1427} & 0.5043 \\
			16 $\times$ 16 & 1.8099 & 0.3867 & 1.9173 & 0.4298 &  \textbf{1.5630} & 0.1780 \\
			32 $\times$ 32 & 1.4536 & 0.1653  & 1.5283 & 0.1484 & \textbf{1.2806} & 0.0467 \\
            64 $\times$ 64 & 1.3270 & 0.1982 & 1.3937 & 0.1268 & \textbf{1.1116} & 0.0196 \\
            128 $\times$ 128 & 1.1350 & 0.0272 & 1.2438 & 0.0496 &  \textbf{1.1092} & 0.0492 \\
			\hline
		\end{tabular}
\label{table:4}
}
\end{table}

\section{Conclusion}
The new lossless compression scheme proposed in this paper is implemented in two stages - the zipper transformation stage and the Huffman coding stage. It was shown that the proposed scheme has the capability of reducing the Huffman coding table while improving the compression ratio. From the experimental results, it is shown that the proposed methods outperform the DCT and FWHT-based methods in compressing image data. In other words, the proposed method indeed serves well as an efficient and effective scheme for lossless image compression. Hence we show that using zipper transform based algorithm, a better compression can be achieved. The proposed algorithms also outperform both DCT and FWHT-based algorithms in terms of implementation time.
\bibliographystyle{IEEEtran}
\bibliography{OmanConf}

\begin{thebibliography}{10}
\providecommand{\url}[1]{#1}
\csname url@samestyle\endcsname
\providecommand{\newblock}{\relax}
\providecommand{\bibinfo}[2]{#2}
\providecommand{\BIBentrySTDinterwordspacing}{\spaceskip=0pt\relax}
\providecommand{\BIBentryALTinterwordstretchfactor}{4}
\providecommand{\BIBentryALTinterwordspacing}{\spaceskip=\fontdimen2\font plus
\BIBentryALTinterwordstretchfactor\fontdimen3\font minus
  \fontdimen4\font\relax}
\providecommand{\BIBforeignlanguage}[2]{{%
\expandafter\ifx\csname l@#1\endcsname\relax
\typeout{** WARNING: IEEEtran.bst: No hyphenation pattern has been}%
\typeout{** loaded for the language `#1'. Using the pattern for}%
\typeout{** the default language instead.}%
\else
\language=\csname l@#1\endcsname
\fi
#2}}
\providecommand{\BIBdecl}{\relax}
\BIBdecl

\bibitem{brunello2003lossless}
D.~Brunello, G.~Calvagno, G.~A. Mian, and R.~Rinaldo, ``Lossless compression of
  video using temporal information,'' \emph{Image Processing, IEEE Transactions
  on}, vol.~12, no.~2, pp. 132--139, 2003.

\bibitem{baligar2003high}
V.~P. Baligar, L.~M. Patnaik, and G.~Nagabhushana, ``High compression and low
  order linear predictor for lossless coding of grayscale images,'' \emph{Image
  and Vision Computing}, vol.~21, no.~6, pp. 543--550, 2003.

\bibitem{li2002image}
R.~Y. Li, J.~Kim, and N.~Al-Shamakhi, ``Image compression using transformed
  vector quantization,'' \emph{Image and Vision Computing}, vol.~20, no.~1, pp.
  37--45, 2002.

\bibitem{sriraam20113}
N.~Sriraam and R.~Shyamsunder, ``3-d medical image compression using 3-d
  wavelet coders,'' \emph{Digital Signal Processing}, vol.~21, no.~1, pp.
  100--109, 2011.

\bibitem{srikanth2005contextual}
R.~Srikanth and A.~Ramakrishnan, ``Contextual encoding in uniform and adaptive
  mesh-based lossless compression of mr images,'' \emph{Medical Imaging, IEEE
  Transactions on}, vol.~24, no.~9, pp. 1199--1206, 2005.

\bibitem{xiong2003lossy}
Z.~Xiong, X.~Wu, S.~Cheng, and J.~Hua, ``Lossy-to-lossless compression of
  medical volumetric data using three-dimensional integer wavelet transforms,''
  \emph{Medical Imaging, IEEE Transactions on}, vol.~22, no.~3, pp. 459--470,
  2003.

\bibitem{nguyen2011efficient}
B.~P. Nguyen, C.-K. Chui, S.-H. Ong, and S.~Chang, ``An efficient compression
  scheme for 4-d medical images using hierarchical vector quantization and
  motion compensation,'' \emph{Computers in biology and medicine}, vol.~41,
  no.~9, pp. 843--856, 2011.

\bibitem{miaou2009lossless}
S.-G. Miaou, F.-S. Ke, and S.-C. Chen, ``A lossless compression method for
  medical image sequences using jpeg-ls and interframe coding,''
  \emph{Information Technology in Biomedicine, IEEE Transactions on}, vol.~13,
  no.~5, pp. 818--821, 2009.

\bibitem{tai2000adaptive}
S.-C. Tai, Y.-G. Wu, and C.-W. Lin, ``An adaptive 3-d discrete cosine transform
  coder for medical image compression,'' \emph{Information Technology in
  Biomedicine, IEEE Transactions on}, vol.~4, no.~3, pp. 259--263, 2000.

\bibitem{sunder2005performance}
R.~S. Sunder, C.~Eswaran, and N.~Sriraam, ``Performance evaluation of 3-d
  transforms for medical image compression,'' in \emph{Electro Information
  Technology, 2005 IEEE International Conference on}.\hskip 1em plus 0.5em
  minus 0.4em\relax IEEE, 2005, pp. 6--pp.

\bibitem{patel2015survey}
H.~Patel, U.~Itwala, R.~Rana, and K.~Dangarwala, ``Survey of lossless data
  compression algorithms,'' in \emph{International Journal of Engineering
  Research and Technology}, vol.~4, no. 04 (April-2015).\hskip 1em plus 0.5em
  minus 0.4em\relax ESRSA Publications, 2015.

\bibitem{alzahir2015innovative}
S.~Alzahir and A.~Borici, ``An innovative lossless compression method for
  discrete-color images,'' \emph{Image Processing, IEEE Transactions on},
  vol.~24, no.~1, pp. 44--56, 2015.

\bibitem{campobello2015comparison}
G.~Campobello, O.~Giordano, A.~Segreto, and S.~Serrano, ``Comparison of local
  lossless compression algorithms for wireless sensor networks,'' \emph{Journal
  of Network and Computer Applications}, vol.~47, pp. 23--31, 2015.

\bibitem{Hinton2006reducing}
G.~Hinton and R.~Salakhutdinov, ``Reducing the dimensionality of data with
  neural networks,'' \emph{Science}, vol. 313, no. 5786, pp. 504--507, 2006.

\bibitem{wang2012folded}
J.~Wang, H.~He, and D.~V. Prokhorov, ``A folded neural network autoencoder for
  dimensionality reduction,'' \emph{Procedia Computer Science}, vol.~13, pp.
  120--127, 2012.

\bibitem{hashim2016optimal}
H.~A. Hashim, B.~Ayinde, and M.~Abido, ``Optimal placement of relay nodes in
  wireless sensor network using artificial bee colony algorithm,''
  \emph{Journal of Network and Computer Applications}, vol.~64, pp. 239--248,
  2016.

\bibitem{egorov2015performance}
N.~Egorov, D.~Novikov, and M.~Gilmutdinov, ``Performance analysis of prediction
  methods for lossless image compression,'' in \emph{Intelligent Interactive
  Multimedia Systems and Services}.\hskip 1em plus 0.5em minus 0.4em\relax
  Springer, 2015, pp. 169--178.

\bibitem{li2001edge}
X.~Li and M.~T. Orchard, ``Edge-directed prediction for lossless compression of
  natural images,'' \emph{Image Processing, IEEE Transactions on}, vol.~10,
  no.~6, pp. 813--817, 2001.

\bibitem{zhang2007efficient}
J.~Zhang and G.~Liu, ``An efficient reordering prediction-based lossless
  compression algorithm for hyperspectral images,'' \emph{Geoscience and Remote
  Sensing Letters, IEEE}, vol.~4, no.~2, pp. 283--287, 2007.

\bibitem{wang2005lossless}
H.~Wang, S.~D. Babacan, and K.~Sayood, ``Lossless hyperspectral-image
  compression using context-based conditional average,'' \emph{IEEE
  transactions on geoscience and remote sensing}, vol.~45, no.~12, pp.
  4187--4193, 2007.

\bibitem{clunie2000lossless}
D.~A. Clunie, ``Lossless compression of grayscale medical images: effectiveness
  of traditional and state-of-the-art approaches,'' in \emph{Medical Imaging
  2000}.\hskip 1em plus 0.5em minus 0.4em\relax International Society for
  Optics and Photonics, 2000, pp. 74--84.

\bibitem{memon1996lossless}
N.~D. Memon and K.~Sayood, ``Lossless compression of video sequences,''
  \emph{Communications, IEEE Transactions on}, vol.~44, no.~10, pp. 1340--1345,
  1996.

\bibitem{memon1997interband}
N.~D. Memon, X.~Wu, V.~Sippy, and G.~Miller, ``Interband coding extension of
  the new lossless jpeg standard,'' in \emph{Electronic Imaging'97}.\hskip 1em
  plus 0.5em minus 0.4em\relax International Society for Optics and Photonics,
  1997, pp. 47--58.

\bibitem{yang2000contex}
K.~H. Yang \emph{et~al.}, ``A contex-based predictive coder for lossless and
  near-lossless compression of video,'' in \emph{Image Processing, 2000.
  Proceedings. 2000 International Conference on}, vol.~1.\hskip 1em plus 0.5em
  minus 0.4em\relax IEEE, 2000, pp. 144--147.

\bibitem{ranade2007variation}
A.~Ranade, S.~S. Mahabalarao, and S.~Kale, ``A variation on svd based image
  compression,'' \emph{Image and Vision computing}, vol.~25, no.~6, pp.
  771--777, 2007.

\bibitem{hernandez2016progressive}
M.~Hern{\'a}ndez-Cabronero, I.~Blanes, A.~J. Pinho, M.~W. Marcellin, and
  J.~Serra-Sagrist{\`a}, ``Progressive lossy-to-lossless compression of dna
  microarray images,'' \emph{IEEE Signal Processing Letters}, vol.~23, no.~5,
  pp. 698--702, 2016.

\bibitem{tomar2015lossless}
R.~R.~S. Tomar and K.~Jain, ``Lossless image compression using differential
  pulse code modulation and its application,'' in \emph{Communication Systems
  and Network Technologies (CSNT), 2015 Fifth International Conference
  on}.\hskip 1em plus 0.5em minus 0.4em\relax IEEE, 2015, pp. 543--545.

\bibitem{hu2000new}
Y.-C. Hu and C.-C. Chang, ``A new lossless compression scheme based on huffman
  coding scheme for image compression,'' \emph{Signal Processing: Image
  Communication}, vol.~16, no.~4, pp. 367--372, 2000.

\bibitem{pinho2002online}
A.~J. Pinho, ``An online preprocessing technique for improving the lossless
  compression of images with sparse histograms,'' \emph{IEEE Signal Processing
  Letters}, vol.~9, no.~1, pp. 5--7, 2002.

\bibitem{miaou2005wavelet}
S.-G. Miaou and S.-N. Chao, ``Wavelet-based lossy-to-lossless ecg compression
  in a unified vector quantization framework,'' \emph{Biomedical Engineering,
  IEEE Transactions on}, vol.~52, no.~3, pp. 539--543, 2005.

\bibitem{bruylants2015wavelet}
T.~Bruylants, A.~Munteanu, and P.~Schelkens, ``Wavelet based volumetric medical
  image compression,'' \emph{Signal Processing: Image Communication}, vol.~31,
  pp. 112--133, 2015.

\bibitem{hans2001lossless}
M.~Hans and R.~W. Schafer, ``Lossless compression of digital audio,''
  \emph{Signal Processing Magazine, IEEE}, vol.~18, no.~4, pp. 21--32, 2001.

\bibitem{ayinde2016lossless}
B.~O. Ayinde and A.~H. Desoky, ``Lossless image compression using zipper
  transformation,'' in \emph{Proceedings of the International Conference on
  Image Processing, Computer Vision, and Pattern Recognition (IPCV)}.\hskip 1em
  plus 0.5em minus 0.4em\relax The Steering Committee of The World Congress in
  Computer Science, Computer Engineering and Applied Computing (WorldComp),
  2016, p. 103.

\bibitem{he2003optimal}
C.~He, J.~Dong, Y.~F. Zheng, and Z.~Gao, ``Optimal 3-d coefficient tree
  structure for 3-d wavelet video coding,'' \emph{Circuits and Systems for
  Video Technology, IEEE Transactions on}, vol.~13, no.~10, pp. 961--972, 2003.

\bibitem{verhack2015lossless}
R.~Verhack, L.~Lange, P.~Lambert, R.~Van~de Walle, and T.~Sikora, ``Lossless
  image compression based on kernel least mean squares,'' in \emph{Picture
  Coding Symposium (PCS), 2015}.\hskip 1em plus 0.5em minus 0.4em\relax IEEE,
  2015, pp. 189--193.

\bibitem{herman20003d}
G.~T. Herman and J.~K. Udupa, \emph{3D imaging in medicine}.\hskip 1em plus
  0.5em minus 0.4em\relax CRC Press, 2000.

\bibitem{scan1}
G.~Templeton, ``Mri scan,'' URL \url{scan:
  http://www.extremetech.com/wp-content/uploads/2013/10/scan.jpg}, 2013.

\bibitem{scan2}
T.~Melzer, ``Mri scan 2,'' URL
  \url{www.otago.ac.nz/healthsciences/research/facilities/otago084592.html},
  2016.

\end{thebibliography}
\end{document}